\documentclass{article}

\usepackage{adjustbox}
\usepackage{longtable}

\usepackage{arxiv}

\usepackage{bm}
\usepackage{amssymb,amsthm,mathtools}

\usepackage[utf8]{inputenc} 
\usepackage[T1]{fontenc}    
\usepackage{url}            
\usepackage{booktabs}       
\usepackage{amsfonts}       
\usepackage{nicefrac}       
\usepackage{microtype}      
\usepackage{lipsum}
\usepackage{multirow}
\usepackage{xcolor}
\usepackage{enumitem}
\usepackage{graphicx}
\usepackage{orcidlink}
\usepackage{comment}
\usepackage{array}
\newcolumntype{R}[1]{>{\raggedleft\arraybackslash}p{#1}}
\newcolumntype{L}[1]{>{\raggedright\arraybackslash}p{#1}}

\usepackage[numbers]{natbib}
\usepackage{caption}
\usepackage{amsmath}
\usepackage{rotating}

\usepackage{siunitx}
\usepackage{placeins}

\usepackage{xspace}


\newcommand{\code}[1]{\texttt{#1}}
\newcommand{\auc}{\text{AUC}}
\newcommand{\auprc}{\text{AUPRC}}
\newcommand{\appsec}[1]{Appendix Section~\ref{#1}}
\newcommand{\apptab}[1]{Appendix Table~\ref{#1}}
\newcommand{\appfig}[1]{Appendix Figure~\ref{#1}}

\makeatletter
\renewcommand{\footnotesize}{\@setfontsize\footnotesize{8pt}{10pt}}
\makeatother

\makeatletter
\renewcommand{\scriptsize}{\@setfontsize\scriptsize{7pt}{9pt}}
\makeatother

\definecolor{VUB_blauw}{rgb}{0.1529, 0.2667, 0.5529}
\usepackage{hyperref}       
\hypersetup{
    colorlinks,%
    citecolor=VUB_blauw,%
    filecolor=VUB_blauw,%
    linkcolor=VUB_blauw,%
    urlcolor=VUB_blauw
}

\usepackage{cleveref}

\graphicspath{{../figures/}{../}{./figures/}{./}}

\AddToHook{shipout/background}{%
  \ifnum\value{page}=1
    \pagenumbering{Roman}
    \setcounter{page}{1}
  \fi
  \ifnum\value{page}=2
  \pagenumbering{arabic}
  \setcounter{page}{2}
  \fi
}

\title{Rebuttals Move Peer-Review Scores, \\ but Initial-Review Structure Bounds the Movement}
\runningtitle{Rebuttals Move Peer-Review Scores}

\author{
  Mathieu Louis\textsuperscript{1,2} \\
  \orcidlinkc{0009-0003-0589-4357} \\
  \And
  Tibo Vanleke\textsuperscript{1,2} \\
  \orcidlinkc{0009-0003-2200-7529} \\
  \And
  Vincent Ginis\textsuperscript{1,2,3} \\
  \orcidlinkc{0000-0003-0063-9608} \\
  \And
  Andres Algaba\textsuperscript{1,2} \\
  \orcidlinkc{0000-0002-0532-3066} \\
  \and
  \textsuperscript{1}Data Analytics Lab, Vrije Universiteit Brussel, Pleinlaan 5, 1050 Brussel, Belgium \\
  \textsuperscript{2}imec-SMIT, Vrije Universiteit Brussel, Pleinlaan 9, 1050 Brussels, Belgium \\
  \textsuperscript{3}School of Engineering and Applied Sciences, Harvard University, Cambridge, Massachusetts 02138, USA
}

\begin{document}

\maketitle
\renewcommand{\thefootnote}{}
\footnotetext{\includegraphics[height=1em]{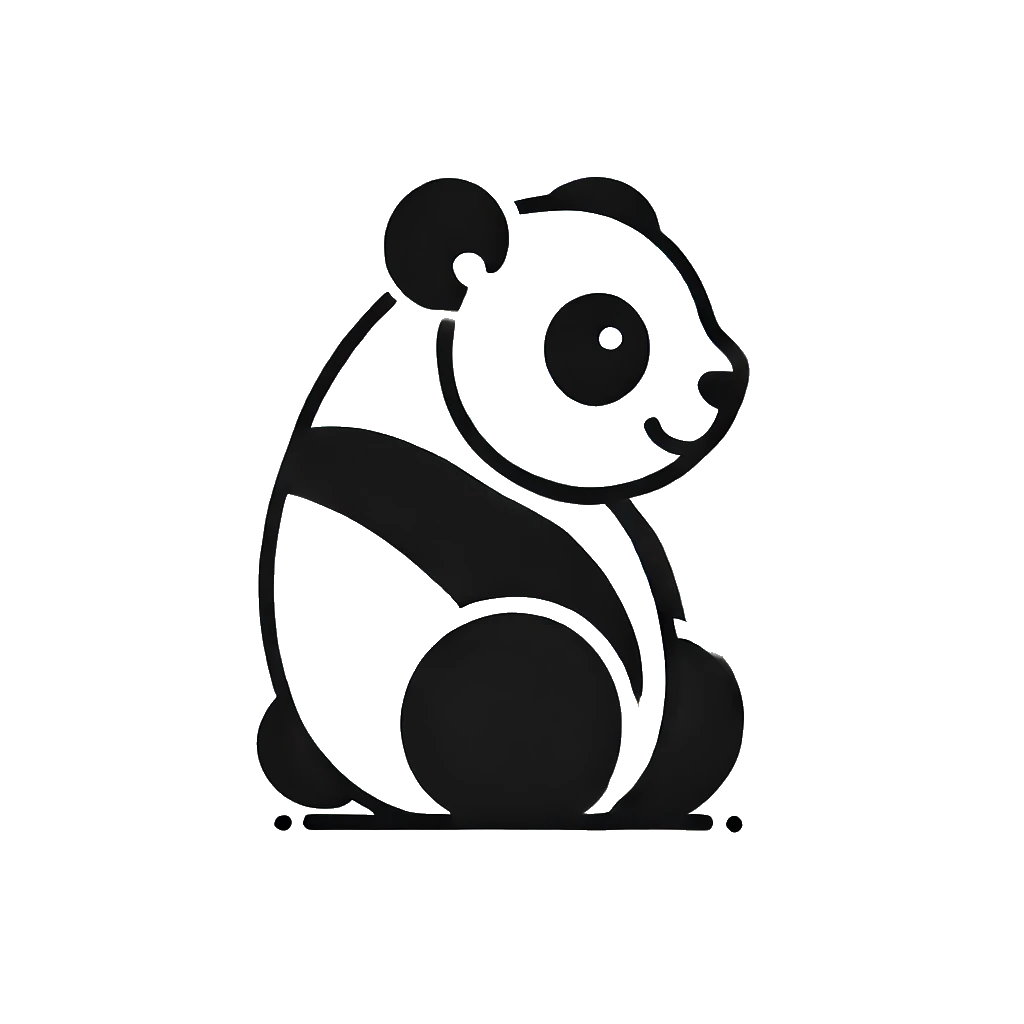} Corresponding author: \href{mailto:mathieu.pierre.louis@vub.be}{mathieu.pierre.louis@vub.be} \\}
\renewcommand{\thefootnote}{\arabic{footnote}}
\thispagestyle{plain}

\begin{abstract}
Author rebuttals are the main post-submission window in peer review, but their effect on reviewer scores remains hard to measure because score updates mix rebuttal content with initial score position, paper-level consensus, reviewer confidence, and discussion dynamics. We study ICLR 2024--2025 using 73,000 reviewer trajectories with externally archived pre- and post-rebuttal scores, and use LLMs only as measurement instruments. Gemini Flash 3.0 predicts implied pre-rebuttal scores from score-stripped review text. The resulting text--score offset predicts later movement, with score-increase rates rising from 8.3\% when text reads below the assigned score to 31.9\% when it reads above. Claude Opus 4.6 induces, and outcome-blinded Gemini Flash 3.0 validates, a 44-feature taxonomy of resolved reviewer--author exchanges, where 23 features replicate across model and held-out year under Bonferroni correction. In the rebuttal-engaged benchmark (n=6,705), initial-review structure already predicts much score movement (AUC=0.747, minimal AUC=0.696), while adding the resolved exchange raises AUC to 0.804. Rebuttals can move scores, but measurable movement is bounded by initial-review structure, and robust exchange signals are mostly rebuttal failure modes.
\end{abstract}


\begin{figure}[!t]
  \centering
  \includegraphics[width=\textwidth]{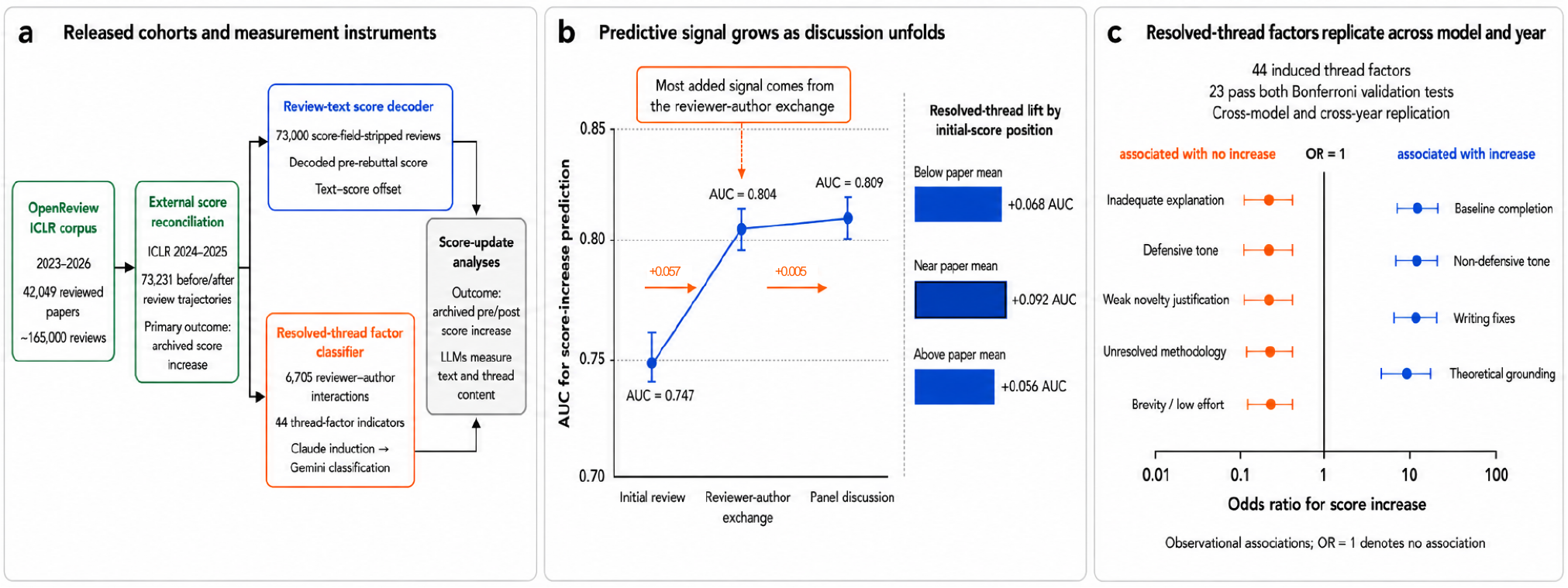}
  \caption{\textbf{ICLR rebuttals expose both structural limits and measurable content signals.} \textbf{a}, Measurement layers and outcome flow: OpenReview records, score predictions from stripped review text, exchange-feature labels, and the external before/after score outcome. \textbf{b}, AUC across the three observation horizons (initial review, resolved exchange, panel discussion). \textbf{c}, Twenty-three of 44 exchange features replicate across an independent validator and a held-out year under Bonferroni.}
  \label{fig:overview}
\end{figure}

\section{Introduction}\label{sec:intro}

Peer review is one of machine learning's central evaluation systems. It converts private assessments of research claims into public selection decisions. Reviewer workload has grown sharply: ICLR submissions rose from \num{4955} in 2023 to \num{19809} in 2026~\cite{papercopilot}. The process is also heterogeneous: review burden is unevenly distributed~\cite{kovanis}, author-identity information can bias recommendations~\cite{tomkins2017reviewer}, and reviewer experience predicts acceptance recommendations in a large finance-journal sample~\cite{hrazdil2026determinants}. Large-scale process audits and consistency experiments find substantial committee-level randomness and review-quality variation~\cite{cortes2014nips,beygelzimer2023consistency,tran2020openreview,goldberg2025peerreviews}. These pressures matter more now that large language models (LLMs) are entering research and reviewing workflows~\cite{musslick2025automating,liao2024llmsresearchtoolslarge,kusumegi2025scientific,si2024can,baek2025researchagent,schmidgall2025agent,boiko2023autonomous,swanson2025virtual,lu2026towards}. If automated systems are to help write, evaluate, or audit reviews, the human evaluation process itself needs reproducible measurement instruments~\cite{ioannidis2005most,open2015estimating,camerer2016evaluating,fortunato2018science}. For NLP, this setting is a test case for outcome-grounded measurement of long, multi-party text: the task is to convert review and rebuttal text into features linked to score changes, not to generate or judge reviews.

The rebuttal phase is the point where that measurement problem becomes operational. A reviewer writes a text review and score, authors respond, discussion follows, and the reviewer may revise the score. Across our ICLR 2024--2025 corpus, $21.0\%$ of matched reviewers raise their score after rebuttal and $1.4\%$ lower it. For authors, this is the main post-submission window where they can address reviewer concerns and influence the reviewer's score. For researchers, however, score movement is not a clean measure of rebuttal quality. It mixes initial score position, paper-level consensus, reviewer confidence, co-reviewer behavior, and author response. The central question is therefore not whether rebuttals matter, but how much update signal is already visible before authors reply, and which parts of the reviewer--author exchange add reproducible information beyond that baseline.

Prior work addresses parts of this question. Peer-review corpora and discourse studies analyse reviews and rebuttals~\cite{kang,hua,dycke2023nlpeer,kuznetsov,kennard,zhang2025re2}. Process studies model score dynamics and discussion in ML peer review~\cite{gao,huang,jung,kargaran,li2025effective,papercopilot2026tracking}. Adjacent work studies the gap between review text and scores, and the detectability of AI-written reviews~\cite{wen2026decoupling,yu2026llmreviewed}. Recent work tests LLM feedback, AI-assisted reviewing, and automated review or rebuttal systems~\cite{liang,thakkar,russo,biswas2026aiassisted,goyal2026scholarpeer,han2026drpg,khatri2026defend,baumann,du2024llmsassist}. A recent expert-annotation study evaluates LLMs as reviewers in their own right: across \num{2960} criticisms from 82 \textit{Nature}-family papers, domain scientists scored a GPT-5.2 agent above the top-rated human reviewer, while documenting recurring failure modes such as miscalibrated severity and losing track of content across long papers~\cite{kim2026aireviewers}. The remaining NLP problem is measurement: how to turn long, multi-speaker review threads into structured variables that are reproducible and outcome-validated, with the LLM acting as feature extractor rather than judge. We use LLMs in that narrower role. They do not judge paper quality, recommend acceptance, or assign the score-update outcome. They extract structured measurements from existing review and discussion text, while the primary label comes from an external before/after score archive~\cite{kargaran}.

\begin{table}[!t]
  \centering
  \caption{\textbf{Four evaluation tasks define the measurement protocol.} The score-increase prediction task uses outputs of the score decoder and exchange-feature measurement as features alongside structural metadata.}
  \label{tab:evaluation-tasks}
  \small
  \setlength{\tabcolsep}{4pt}
  \renewcommand{\arraystretch}{1.1}
  \begin{tabular}{L{2.8cm}L{5cm}L{2.8cm}L{0.7cm}L{3.3cm}}
    \toprule
    Task & Input layer & Target or output & \S\,ref. & Metrics \\
    \midrule
    Engagement contrast & Matched reviews with and without author reply & External score increase & \S\ref{res:engagement-gate} & Increase rates, initial-score structure \\
    Score decoder & Initial review text with score fields stripped & External pre-rebuttal score & \S\ref{res:instrument} & Exact match, within-one-step, Spearman, offset gradient \\
    Exchange-feature measurement & Review, author response, and reviewer follow-up & 44 exchange-feature indicators & \S\ref{res:taxonomy} & Odds ratios, CIs, permutation tests \\
    Score-increase prediction & H0 initial review, H1 reviewer-author exchange (uses score decoder and feature indicators), H2 panel discussion & External post-score $>$ pre-score & \S\ref{sec:rebuttal-impact} & ROC AUC, AUPRC, calibration \\
    \bottomrule
  \end{tabular}
  \renewcommand{\arraystretch}{1.0}
\end{table}
\begin{table}[!b]
  \centering
  \caption{\textbf{The prediction benchmark uses the part of the matched score-update record where the reviewer-author exchange can be characterized.} Base rate is 45.6$\%$, against $34.4\%$ for the broader bidirectional-engagement cohort. Acceptance counts Oral, Spotlight, and Poster.}
  \label{tab:cohort-flow-benchmark}
  \small
  \setlength{\tabcolsep}{5pt}
  \begin{tabular}{L{6cm}R{1.7cm}R{1.7cm}R{1.7cm}R{1.7cm}R{1.9cm}}
    \toprule
    Cohort step & Rows & Score increases & Increase rate & Mean initial score & Paper acceptance rate \\
    \midrule
    Externally matched review trajectories & \num{73231} & \num{15394} & 21.0\% & 4.81 & 31.7\% \\
    Author replied to the review & \num{55241} & \num{14903} & 27.0\% & 5.09 & 40.9\% \\
    Author reply plus non-author follow-up & \num{35498} & \num{12224} & 34.4\% & 5.07 & 45.1\% \\
    Exchange labels and external outcome available & \num{6730} & \num{3069} & 45.6\% & 5.10 & 49.7\% \\
    Three-stage prediction benchmark & \num{6705} & \num{3056} & 45.6\% & 5.10 & 49.7\% \\
    \bottomrule
  \end{tabular}
\end{table}

We present an LLM-based measurement protocol for score-update dynamics in ICLR peer review. The protocol has three components. First, a score decoder predicts the implied score from review text with explicit scores removed. Second, an induce-then-validate taxonomy converts resolved reviewer--author exchanges into 44 content features, using one LLM for extraction and an independent LLM for outcome-blinded classification. Third, a three-horizon prediction benchmark asks how much signal is visible before the author response, after the resolved exchange, and after the panel discussion. The main analyses use ICLR 2024--2025, where external before/after scores are available for approximately \num{73000} matched reviews, of which \num{6705} form the three-stage prediction benchmark with parsed exchange features and score-decoder features (\Cref{tab:cohort-flow-benchmark} traces the full review funnel). In tables and appendix specifications, we use compact labels for the same stages: H0 for the initial review, H1 for the reviewer--author exchange, and H2 for co-reviewer movement during the panel discussion. \Cref{tab:evaluation-tasks} states the four evaluation tasks supported by the measurement layers. We also release raw ICLR 2023--2026 OpenReview records, normalized tables, split definitions, prompts, analysis code, and responsible-use notes.

First, score-stripped review text contains a measurable residual: Gemini Flash 3.0 recovers the exact pre-rebuttal score in 40.7\% of reviews, but the more useful signal is the text--score offset, whose score-increase gradient runs from 8.3\% below the assigned score to 31.9\% above it. Second, Claude Opus 4.6 with extended thinking enabled, followed by outcome-blinded Gemini Flash 3.0 classification, produces 44 exchange features, 23 of which replicate across an independent validator and a held-out year under Bonferroni correction. Third, much recoverable score-increase signal is already present before the author writes: a provenance-minimal initial-review block reaches $\auc=0.696$, and the full initial-review block reaches $0.747$. Adding the resolved reviewer--author exchange raises this to $0.804$, with H2 panel discussion adding only $0.005$. Finally, reviews without an author reply increase only 2.7\% of the time, compared with 27.0\% among reviews with a reply. In practical terms, author participation is a strong selection boundary for score improvement, but once authors participate, reviewer variability is already substantially structured before the rebuttal appears.

\Cref{fig:overview} summarizes the measurement layers and prediction stages. The results follow the title claim: author response gates which reviews can move at all, and within the engaged cohort, the resolved reviewer--author exchange adds a bounded but reproducible increment beyond what is already visible in the initial review.

\section{Author participation is a selection boundary, not a treatment}\label{res:engagement-gate}
\begin{figure}[!b]
  \centering
  \begin{minipage}[t]{0.48\textwidth}
    \centering
    \includegraphics[width=\linewidth]{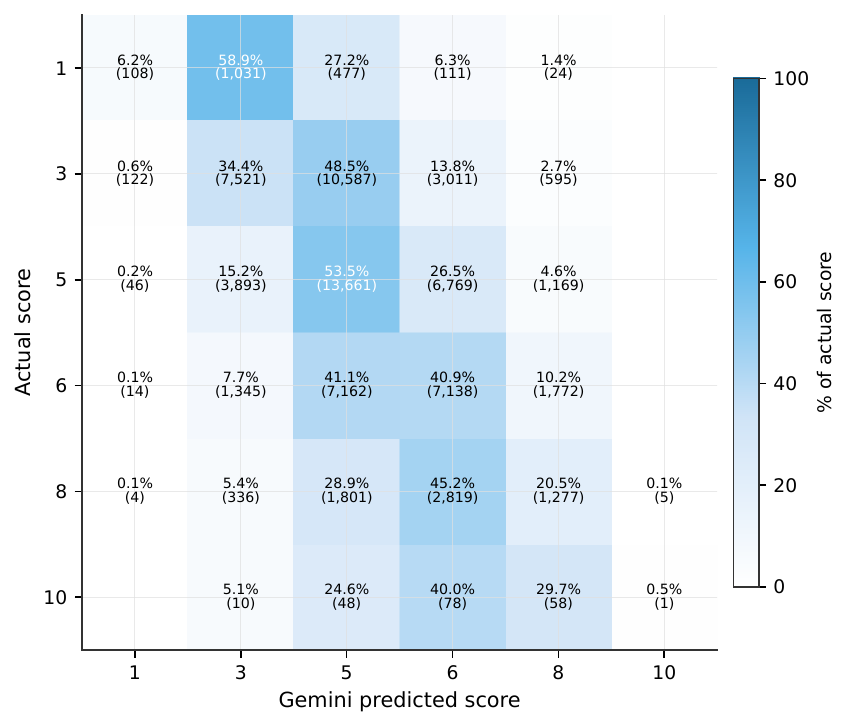}
    \vspace{0.2em}

    {\footnotesize \textbf{(a)} Score-decoding accuracy}
  \end{minipage}\hfill
  \begin{minipage}[t]{0.48\textwidth}
    \centering
    \includegraphics[width=\linewidth]{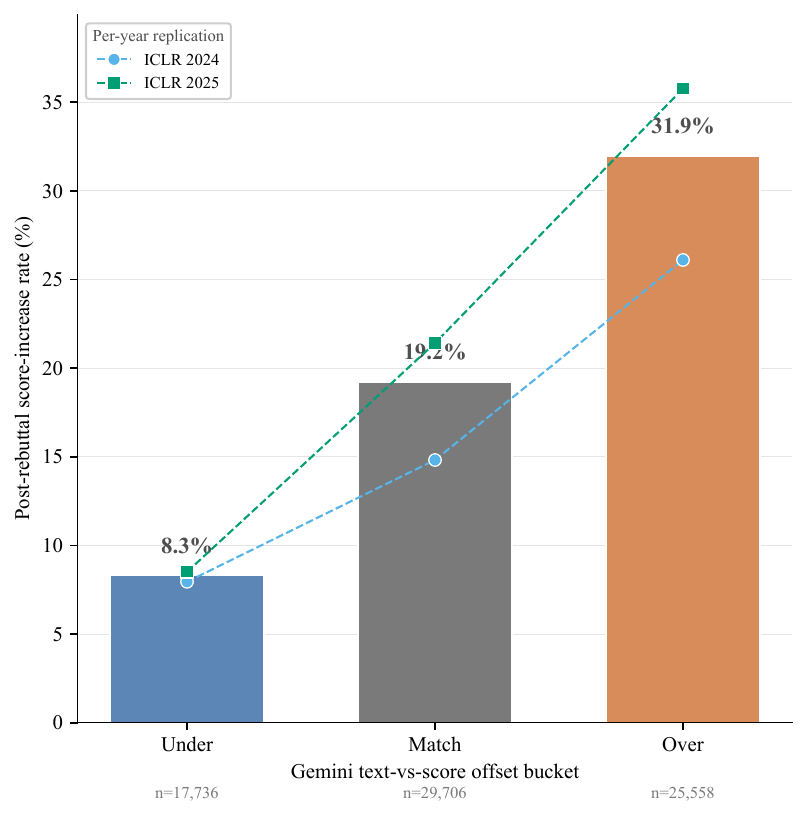}
    \vspace{0.2em}

    {\footnotesize \textbf{(b)} Text--score offset gradient}
  \end{minipage}
  \caption{\textbf{Score-stripped review text supplies a checkable measurement channel.} \textbf{a}, Row-normalized confusion matrix of Gemini-decoded vs.\ external pre-rebuttal scores ($n=\num{73000}$). \textbf{b}, The decoded-minus-assigned offset predicts later movement: score-increase rates rise from $8.3\%$ (text below assigned) to $31.9\%$ (text above). Bars pool years, markers show 2024 and 2025 separately.}
  \label{fig:f2-instrument}
\end{figure}
\Cref{tab:cohort-flow-benchmark} and \apptab{tab:engagement-contrast} show the first boundary condition. Among approximately \num{73000} externally matched ICLR 2024--2025 review trajectories, \num{17990} have no direct author reply to the review-rooted thread. Only 491 of those no-reply reviews increase, an increase rate of 2.7\%. Reviews with at least one author reply increase ten times more often, at 27.0\% (\num{14903} of \num{55241}). The contrast is observational: the two cohorts differ on starting position as well as on engagement. No-reply reviews start at a lower initial score (mean 4.0 versus 5.1) and below the paper-level initial mean (deviation $-0.32$ versus $+0.10$), and authors may skip rebuttal for many reasons beyond score.

The prediction benchmark is drawn from the engaged side of this boundary. It retains reviewer--paper rows with author reply, parsed exchange labels, Gemini score-decoder features, and external before/after scores, yielding \num{6705} rows with a 45.6\% score-increase base rate. This enrichment is intentional. It lets us ask what remains predictable once the author participates and the reviewer--author exchange can be measured. The initial-review, exchange, and panel-discussion horizons below should therefore be read as a decomposition inside the rebuttal-engaged cohort, not as a population-level model of every ICLR review. Because no-reply reviews start systematically lower (mean 4.0 versus 5.1) and below the paper mean, this contrast is selection, not a rebuttal effect. We report it to motivate conditioning all subsequent analyses on the engaged cohort, not as an estimate of participation's causal effect.

\section{Review text provides a calibration signal}\label{res:instrument}

\Cref{fig:f2-instrument} gives the two diagnostics that make the score decoder useful: score recovery from stripped archived review text and the offset gradient linking text--score disagreement to later score movement. We first test whether score-stripped archived review text contains a recoverable numerical judgment. Before sending a review to Gemini, we remove every numeric scoring field the reviewer filled in, including \code{soundness}, \code{presentation}, \code{contribution}, \code{rating}, \code{recommendation}, and self-reported \code{confidence}. The model then predicts the score implied by the remaining text. We compare this prediction against the external pre-rebuttal score, reconciled in \apptab{tab:external-gt}, and measure the text--score offset in ordinal step positions on the ICLR scale $\{1, 3, 5, 6, 8, 10\}$.

Across \num{73000} reviews, Gemini's predictions span the full score scale rather than collapsing to the modal score. Exact agreement is $40.7\%$, within-one-step agreement is $87.6\%$, and Spearman rank correlation is $\rho=0.40$ (\Cref{fig:f2-instrument} and \appsec{app:gemini-pred}). The within-one number is descriptive rather than load-bearing: a constant modal-score predictor reaches $88.8\%$ within one step on this score distribution. The useful signal is disagreement between the written judgment and the assigned score. Gemini reads $24.3\%$ of reviews below the assigned score and $35.0\%$ above it, with a mean offset of $+0.14$ steps.

The offset predicts later updating. Post-rebuttal score-increase rates rise from $8.3\%$ when the text reads below the assigned score to $19.2\%$ at agreement and $31.9\%$ when it reads above, with full counts in \apptab{tab:offset-gradient}. Decreases show the mirror pattern ($3.3\%$, $1.0\%$, and $0.4\%$, against a $1.4\%$ baseline). To rule out that the offset is just a proxy for other Gemini-derived text ratings, we add Gemini's per-review engagement, sentiment, and tone labels as covariates to a logistic regression. The offset coefficient drops by
$23.5\%$, so about three-quarters of its predictive value remains after controlling for these auxiliary signals (\appsec{app:text-score-gap}).

\begin{figure}[!t]
  \centering
  \includegraphics[width=\textwidth,height=0.44\textheight,keepaspectratio]{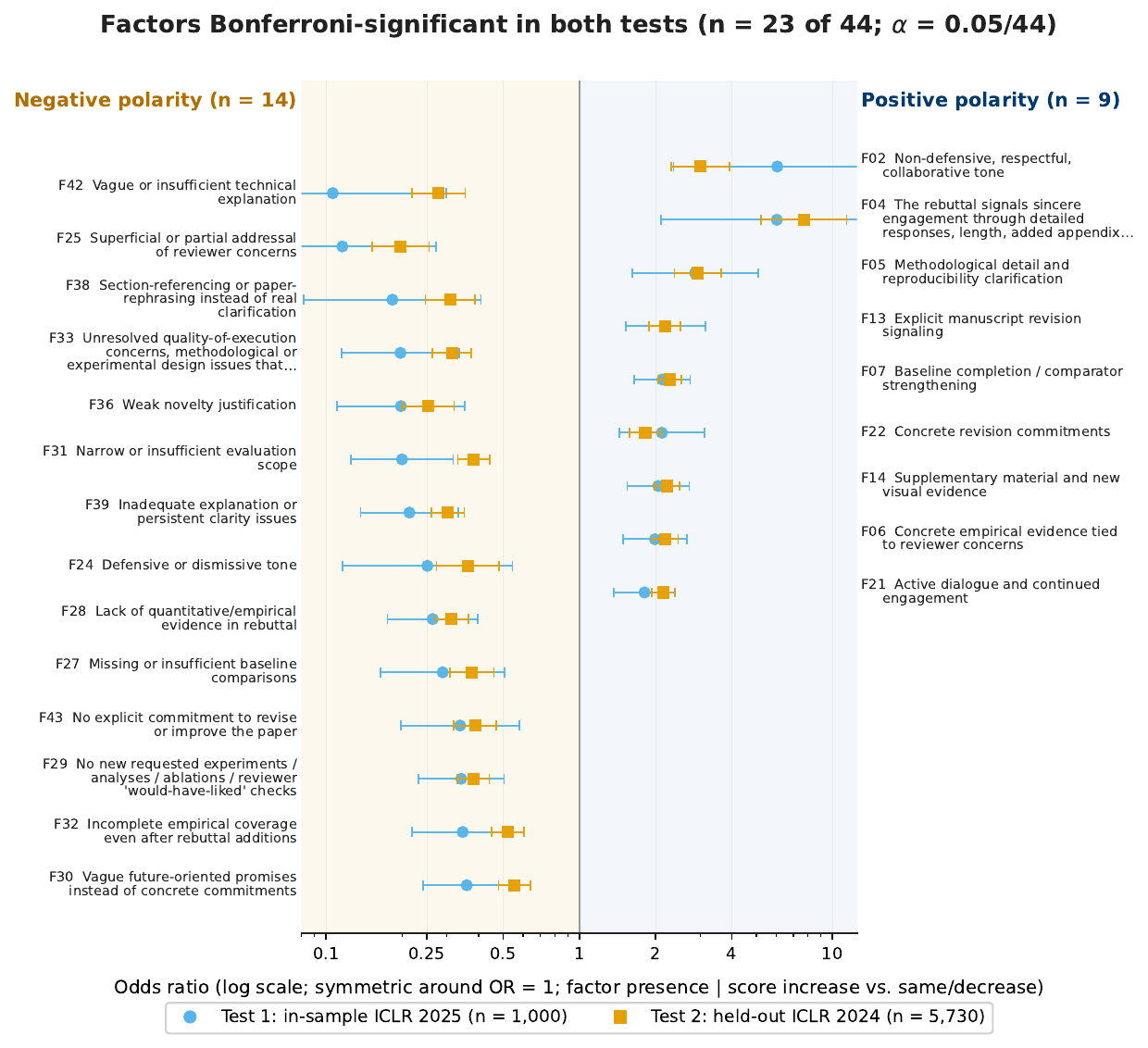}
  \caption{\textbf{The Bonferroni-core exchange features replicate across an independent validator and a held-out year.} Each row is one of the 23 induced features that cross Bonferroni in both tests, plotted as an odds ratio for score increase versus same/decrease (log scale). Test 1: ICLR 2025 induction sample ($n=\num{1000}$, blue circles). Test 2: held-out ICLR 2024 interactions ($n=\num{5730}$, orange squares). The full 44-feature forest is shown in \appfig{fig:f3-categories-full-app}.}
  \label{fig:f3-categories}
\end{figure}

\section{An induce-and-validate LLM workflow yields a reproducible exchange taxonomy}\label{res:taxonomy}

\Cref{fig:f3-categories} is the cross-model and cross-year validation screen for the 44-feature taxonomy. The input is a long, multi-speaker text object: the original review, the author rebuttal, and any subsequent reviewer and author follow-ups. Rather than prompt one model to label a pre-existing codebook, we separate feature discovery from feature application. Claude Opus 4.6 with extended thinking enabled induced candidate features from \num{1000} ICLR 2025 interactions stratified by initial-score tier and outcome (\code{increase} versus \code{same\_decrease}). Five repeated runs are aggregated into 44 features with declared positive or negative polarity. Two memory-control manipulations, one swapping score tiers and one silently swapping outcome labels, show that the features follow the interaction text rather than the labels supplied to the model (\appsec{app:rebuttal-coding}).

An independent Gemini Flash 3.0 classifier then applies the fixed 44-feature list under an outcome-blinded protocol with shuffled feature identifiers. Test 1 reuses the \num{1000} ICLR 2025 induction interactions to ask whether a second model can recover the same features from the same text. Test 2 applies the same protocol to the full disjoint ICLR 2024 cohort of \num{5730} interactions. For each feature we estimate an odds ratio with log-scale confidence interval and a two-sided 2000-permutation test. Test 2 also includes a paper-block permutation robustness check. Bonferroni correction sets the conservative threshold at $\alpha=0.05/44=0.0011$.

Test 1 preserves 43 of 44 declared polarities, and Test 2 preserves all 44. Twenty-three features cross Bonferroni in both tests (9 positive, 14 negative), forming the conservative core of the taxonomy (\Cref{fig:f3-categories} and \apptab{tab:taxonomy-core}). The full 44-feature forest is shown in \appfig{fig:f3-categories-full-app}. Under Benjamini-Hochberg $q=0.05$, the doubly significant set grows to 31 of 44, without removing any Bonferroni-core feature (\apptab{tab:mtcorrection}). The validated positives include sincere engagement, baseline completion, methodological detail, concrete empirical evidence, and active dialogue. The validated negatives include defensive tone, superficial concern addressal, weak novelty justification, missing baselines, vague future promises, and section-referencing instead of clarification. We treat the taxonomy as a measurement instrument for reviewer--author exchanges. The features are not a ground-truth ontology of rebuttal quality or reviewer merit.

\section{Exchange features retain signal after structural controls}\label{sec:horizons}

\begin{figure}[!htb]
  \centering
  \includegraphics[width=\textwidth,height=0.48\textheight,keepaspectratio]{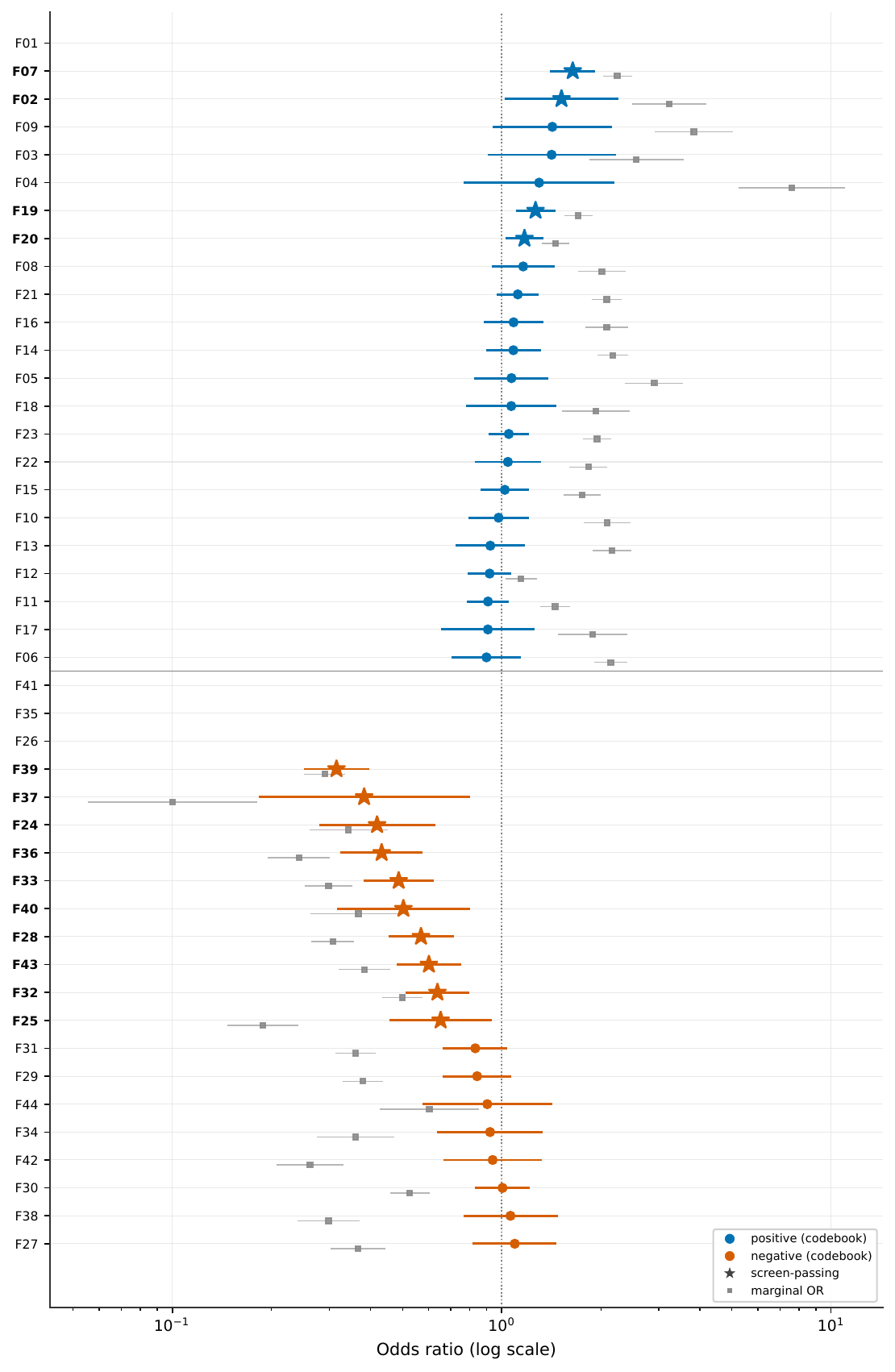}
  \caption{\textbf{Only a subset of exchange features retains signal after initial-review controls.} Positive-polarity features (blue) and negative-polarity features (orange) on a shared log-OR axis. Large markers: polarity-isolated conditional OR after controlling for initial-review structure. Small grey squares: univariate marginal OR. Stars mark the 14 features that pass the asymmetric screen (4 positives, 10 negatives). Screen definition in \Cref{sec:horizons}.}
  \label{fig:recforest}
\end{figure}

\Cref{fig:recforest} asks which exchange features retain conditional signal after controlling for initial-review structure. To separate exchange content from that structure, we fit a paper-clustered logistic regression on the combined ICLR 2024--2025 exchange prediction benchmark ($n=\num{6705}$, score-increase base rate $45.6\%$). The model includes the 44 feature indicators, ten initial-review structural controls, and four interaction-side aggregates drawn from prior analyses of peer-review dynamics~\cite{huang,jung,kargaran}. Full feature definitions, leakage checks, provenance notes, and robustness fits are in \appsec{app:horizons}.

Positive and negative features tend to be anti-correlated within the same exchange, so a single joint coefficient can understate both sides. We therefore use an asymmetric screen: a feature must survive both a marginal fit and a polarity-isolated joint fit, with the 95\% CI excluding $1$ in both. Four of 22 testable positives pass this screen: baseline completion ($F_{07}$, OR $=1.64$ [$1.42$, $1.89$]), non-defensive tone ($F_{02}$, $1.51$), writing fixes ($F_{19}$, $1.27$), and theoretical grounding ($F_{20}$, $1.18$). Ten of 18 testable negatives pass, led by inadequate explanation ($F_{39}$, OR $=0.31$ [$0.25$, $0.39$]), low-effort rebuttal ($F_{37}$, $0.38$), defensive tone ($F_{24}$), and weak novelty justification ($F_{36}$). The declared direction is preserved more often than chance for positives (16/22, $p=0.026$) and negatives (15/18, $p=0.0038$, one-sided binomial sign tests, with details in \appsec{app:horizons}).

The largest coefficients come from structure rather than rhetoric. Initial-score position has the strongest association with later increase (per 1 SD, OR $=0.141$ [$0.122$, $0.163$]), followed by paper-level initial mean (OR $=1.89$) and the \code{contribution} sub-score (OR $=1.70$). The text--score offset remains associated after these controls (OR $=1.15$ [$1.06$, $1.26$]). Thread depth also contributes (OR $=1.24$, $p=1.0\times10^{-8}$), while rebuttal length, number of author replies, and first-reply latency are flat in the joint model. The minimum-detectable-effect calculation in \apptab{tab:null-triplet-mde} shows 80\% power for per-SD ORs of roughly $1.20$ to $1.22$ on those null features, so it rules out moderate effects but not small ones.

The same pattern appears when we stratify reviewers by their deviation from the paper-level initial mean. Score increases occur in $62.8\%$ of below-mean reviews ($n=\num{2602}$), $46.0\%$ near the mean ($n=\num{2302}$), and $20.2\%$ above it ($n=\num{1801}$) (\appfig{fig:movability-strata-app}). $F_{07}$ baseline completion and $F_{39}$ inadequate explanation keep their direction in all three strata, while weaker screened features have wider intervals (\apptab{tab:movability-feature-coefficients}). These odds ratios are associations in the observed exchanges, not estimates of what would happen if an author inserted one feature into an unchanged rebuttal.

\section{Most score-increase signal is visible before author response}\label{sec:rebuttal-impact}

\begin{figure}[!t]
  \centering
  \includegraphics[width=\textwidth]{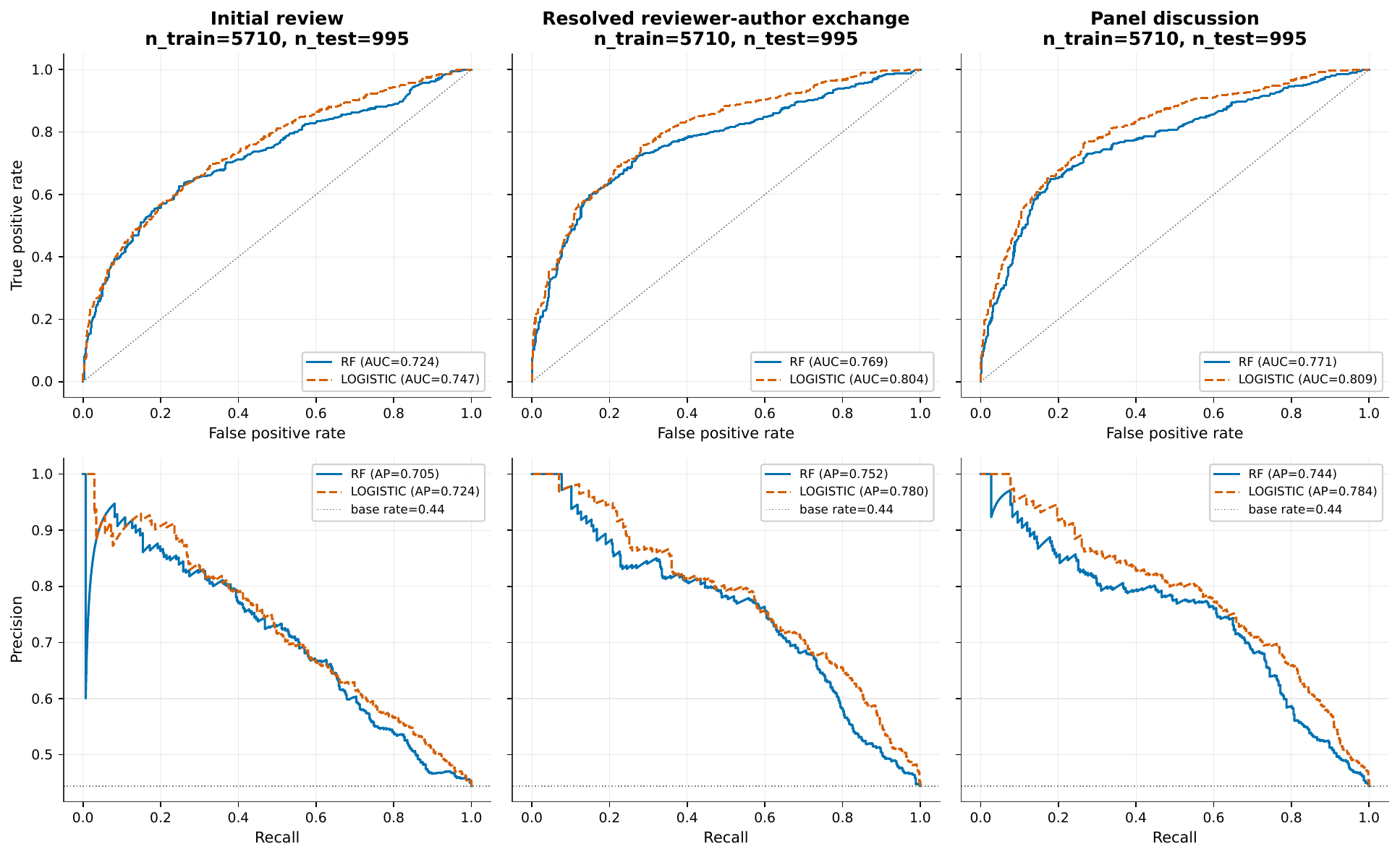}
  \caption{\textbf{The resolved reviewer-author exchange adds a bounded predictive gain.} ROC (top) and precision-recall (bottom) curves for random forest (solid blue) and L2-regularised logistic regression (dashed orange) on the temporal split (train ICLR 2024 $n=\num{5710}$, test stratified ICLR 2025 $n=995$). Columns (left to right): H0 initial review, H1 resolved exchange, H2 panel discussion. Stratum lifts in \apptab{tab:horizon-metrics}.}
  \label{fig:content-lift}
\end{figure}

\Cref{fig:content-lift} separates prediction at three points in the review process. The outcome is a binary reviewer-paper indicator equal to $1$ when the external post-rebuttal rating exceeds the pre-rebuttal rating. Decreases are rare ($1.4\%$ among score-decoding reviews) and are included in the negative class. The benchmark cohort is restricted to reviews with at least one author reply and parsed exchange features, because there is no exchange measurement otherwise. The initial-review benchmark is therefore a baseline within the exchange prediction cohort only.

We compare three nested feature blocks that reveal the prediction process step by step. H0 (initial review) captures everything visible before the author responds. H1 (resolved exchange) adds the reviewer--author rebuttal thread. H2 (panel discussion) adds co-reviewer movement features from the panel period. We report area under the ROC curve ($\auc$) and area under the precision-recall curve ($\auprc$) under both a temporal split (train ICLR 2024, test stratified ICLR 2025 slice) and paper-clustered $\texttt{GroupKFold}(5)$ (\appsec{app:horizons}).

H0 uses the focal reviewer's initial score, sub-scores, confidence, review lengths, paper-level score statistics, deviation from the paper mean, and the text--score offset. The full block reaches $\auc=0.747$ [$0.72$, $0.77$] and $\auprc=0.724$ [$0.68$, $0.76$] on the temporal test, with deviation from the paper mean, initial-score position, and text--score offset carrying much of the signal. A stricter H0-minimal ablation keeps only external pre-score context and immutable timing, excluding mutable OpenReview review-form text and Gemini-derived review-text measurements. H0-minimal reaches $\auc=0.696$ [$0.66$, $0.73$] and $\auprc=0.632$ [$0.59$, $0.68$] (\apptab{tab:pre-no-score}).

H1 adds rebuttal length, reply count, first-reply latency, thread depth, and the 44 feature indicators from the reviewer--author exchange. Adding these raises performance to $\auc=0.804$ [$0.78$, $0.83$] and $\auprc=0.780$ [$0.75$, $0.81$]. The temporal gain is $\Delta\auc=+0.057$ and replicates under paper-clustered cross-validation at $+0.069$. The H1 block characterises the full resolved exchange, not the author's first response alone: the feature classifier reads the original review, the author rebuttal, and any subsequent reviewer and author follow-ups, while the score-update outcome comes from external archived data, so the LLM remains separate from the outcome. The exchange lift is largest near the paper-level initial mean ($+0.092$ AUC), smaller below the mean ($+0.068$), and smallest above it ($+0.056$).

H2 adds co-reviewer movement features computed over the other reviewers of the same paper. These four features (defined in \appsec{app:horizons}) are post-discussion process observables, not inputs for live reviewer-level prediction. Adding them lifts performance by only $+0.005$ on the temporal split and $+0.009$ in cross-validation. The near-zero lift confirms that the predictive ceiling is reached by H1 rather than indicating a missing signal at H2.

\section{Discussion}\label{sec:discussion}

Overall, the results give a calibrated answer to the author-facing question that motivates the study. Author participation is a strong selection boundary for reviewer movement: in the broader matched corpus, reviews with no author reply ($n=\num{17990}$) increase only $2.7\%$ of the time, compared with $27.0\%$ for reviews with a reply (\apptab{tab:engagement-contrast}). However, participation is a threshold condition rather than a guarantee. Among reviews that enter the rebuttal-engaged benchmark, much recoverable variability is already visible before the author writes anything, through initial score position, paper-level consensus, timing, review text, and the text--score offset.

Because the ICLR scale is bounded and ordinal, initial-score position is partly a headroom variable. It is the dominant predictor of later increase (per 1 SD, OR $=0.141$ [$0.122$, $0.163$]), but a review near the ceiling has little room to rise: increase rates fall from $62.8\%$ below the paper-level initial mean to $20.2\%$ above it. The H0 signal should be read as structural movability, not as evidence that the reviewer privately decided before the rebuttal.

The reviewer--author exchange still matters. Adding H1 exchange features lifts temporal-split AUC from $0.747$ to $0.804$, and the gain replicates under paper-clustered cross-validation. The temporal test set is a stratified, rare-cell-enriched ICLR 2025 slice ($n=995$), not a population draw. The paper-clustered \texttt{GroupKFold} over both years (logistic H0/H1/H2 AUC $=0.779$ / $0.848$ / $0.857$, \apptab{tab:horizon-metrics}) is the less position-sensitive estimate. We treat the temporal split as the conservative cross-year check and report both in-text. The largest lift appears for reviewers near the paper-level initial mean, where the reviewer has room to move but is not already far from the paper's consensus. Within that bounded lever, the exchange screen is markedly asymmetric: 10 of 18 testable negatives survive, but only 4 of 22 testable positives. Measurable failure modes are robust: inadequate explanation ($F_{39}$), low-effort rebuttal ($F_{37}$), defensive tone ($F_{24}$), and weak novelty justification ($F_{36}$). The surviving positives (baseline completion, non-defensive tone, writing fixes, theoretical grounding) are few and weak. What the exchange detects is mostly the absence of mistakes, not the presence of persuasion. These are associations in resolved interactions, not causal instructions for inserting phrases into a rebuttal.

Methodologically, the study targets a broader NLP problem: outcome-grounded annotation of long, high-stakes interaction text. The primary outcome is an external before/after score trajectory, not an LLM preference judgment. Gemini decodes a score from score-stripped review text, exposing a within-review offset associated with later movement. Claude Opus 4.6 with extended thinking enabled induces exchange features from stratified interactions, and Gemini validates the fixed taxonomy under outcome-blinded, shuffled-feature classification in a held-out year. This cross-model and cross-year design is stronger than a single-model annotation pass and creates a reusable measurement layer for later audits.

\section{Limitations}\label{sec:limitations}

The study is bounded by scope, identification, and reproducibility. Score-update labels are available only for ICLR 2024--2025, and the rating scale is ordinal with uneven steps, so cross-venue generalisation is not directly tested. The ICLR 2025 feature-validation cohort is stratified and rare-cell-enriched rather than population-representative, which means in-sample feature-validation tests cannot be reweighted to a population estimate without revisiting that stratification. The prediction benchmark conditions on reviews where the author replied ($n=\num{6705}$). Results in \Cref{sec:rebuttal-impact} therefore apply within the engaged cohort rather than to every matched review. Paper-clustered standard errors reduce but do not remove confounding from unobserved paper-level or reviewer-level variables. The 44-feature taxonomy was validated cross-model and cross-year against the archived score outcome but not against a human-coded gold standard, so the feature labels reflect what two independent LLMs agree to extract from text rather than ground-truth content judgements. The induce-and-validate pipeline relies on closed-source LLM APIs (Claude Opus 4.6 and Gemini Flash 3.0). Full re-derivation requires paid API access, though we redistribute consolidated batch outputs so the released tables and figures can be rebuilt without further LLM calls. Both corpus years sit inside the era of AI-assisted reviewing and rebuttal-writing, and ICLR 2025 overlaps Thakkar et al.'s randomized LLM-feedback intervention~\cite{thakkar} with non-public treatment assignment. Our extracted features may therefore partly reflect AI-generated rebuttal style, and observed movement may partly reflect AI-assist effects. The external before/after archive insulates the outcome label from this contamination, but the measured substrate is itself being transformed by systems adjacent to our instruments. We flag this as a first-order threat to the feature taxonomy's interpretation, not a resolved one.

\section{Ethical Considerations}\label{sec:ethics}

Peer review affects who enters the machine-learning literature and whose claims become visible. A reusable record of score updating can support process audits, benchmark evaluation tools against real reviewer behavior, and make claims about rebuttal dynamics easier to check. The same measurements also create misuse risks: participant profiling, deanonymization attempts, live conference monitoring, and systems that optimize author responses for score movement rather than scientific clarity. The dataset is intended for audits, measurement research, and benchmarking of peer-review evaluation protocols. It should not be used for reviewer surveillance, live conference decisions, participant-level targeting, or causal recipes for manipulating scores. The next empirical step is cross-venue replication on NeurIPS, ICML, ACL, and less-anonymised conferences, followed by prospective studies that test whether measured interaction features correspond to actual process changes.

\section*{Acknowledgements}
This research was supported by funding from the Flemish Government under the ``Onderzoeksprogramma Artifici\"ele Intelligentie (AI) Vlaanderen'' program.
Andres Algaba acknowledges support from the Francqui Foundation (Belgium) through a Francqui Start-Up Grant and a fellowship from the Research Foundation Flanders (FWO) under Grant No.1286924N.
Vincent Ginis acknowledges support from Research Foundation Flanders under Grant No.G032822N and G0K9322N.

\section*{Author contributions}
All authors designed the study, developed the measurement protocol, conducted the analysis, and wrote the manuscript.

\section*{Data and code availability}
The analysis code is available on GitHub at \url{https://github.com/Integrated-Intelligence-Lab/iclr-rebuttal-analysis}, and the data are archived on Zenodo at \url{https://doi.org/10.5281/zenodo.20772020}, with a convenience mirror on Hugging Face at \url{https://huggingface.co/datasets/MlouisBE/iclr-rebuttal-analysis}. Together these provide raw OpenReview records for ICLR 2023--2026, normalized tables, the external before/after score reconciliation, Gemini score predictions from stripped review text, 44-feature labels for reviewer--author exchanges, temporal and paper-clustered split definitions, prompt templates, analysis scripts, consolidated batch outputs, and responsible-use notes. The externally archived pre- and post-rebuttal ICLR 2024--2025 reviewer scores from Kargaran et al.~\cite{kargaran} are available at \url{https://github.com/papercopilot/iclr-insights}. Full release contents and licensing are documented in \appsec{app:data-code-availability}.

\clearpage
\bibliographystyle{unsrt}
\bibliography{references}

\clearpage
\appendix
\makeatletter
\@addtoreset{equation}{section}
\makeatother
\renewcommand{\theequation}{\arabic{equation}}
\renewcommand{\theHequation}{app.\thesection.\arabic{equation}}

\setcounter{figure}{0}
\renewcommand{\thefigure}{A\arabic{figure}}
\renewcommand{\theHfigure}{appendix.figure.\arabic{figure}}
\setcounter{table}{0}
\renewcommand{\thetable}{A\arabic{table}}
\renewcommand{\theHtable}{appendix.table.\arabic{table}}

\renewcommand{\tablename}{Appendix Table}
\renewcommand{\figurename}{Appendix Figure}

\section{Data and Code Availability}\label{app:data-code-availability}

The analysis code is available on GitHub at \url{https://github.com/Integrated-Intelligence-Lab/iclr-rebuttal-analysis}, and the data are archived on Zenodo at \url{https://doi.org/10.5281/zenodo.20772020} (mirrored on Hugging Face at \url{https://huggingface.co/datasets/MlouisBE/iclr-rebuttal-analysis}). The code repository bundles the analysis-ready parquet tables and the released figures and tables, so the headline analyses rebuild from a clone. The Zenodo record holds the full data: raw OpenReview records for ICLR 2023--2026, normalized JSON, external before/after score reconciliation, Gemini score predictions from stripped review text, 44-feature labels for reviewer--author exchanges, temporal and paper-clustered split definitions, prompt templates, consolidated batch outputs, and responsible-use notes. The Zenodo record includes a validated Croissant metadata file and dataset documentation following datasheet, data-card, and Croissant principles~\cite{gebru2021datasheets,pushkarna2022datacards,akhtar2024croissant}.

The underlying OpenReview review and discussion records are available from OpenReview under its published terms~\cite{openreview-terms}. Review comments, discussion messages, configuration records, and LLM-derived per-review labels are redistributed under CC-BY-4.0, matching OpenReview's comment licence. Submission and venue metadata are redistributed under CC0 where OpenReview classifies them as such. Downloaded submission PDFs and primary submission body text are not redistributed. Public submission metadata such as titles, abstracts, author lists, and PDF URLs is retained where OpenReview exposes it.

The externally archived pre- and post-rebuttal ICLR 2024--2025 reviewer scores from Kargaran et al.~\cite{kargaran} are available at \url{https://github.com/papercopilot/iclr-insights}.

The released text is public but sensitive. It contains review and discussion messages, pseudonymous reviewer identities, and possible self-disclosures in comments. Intended uses include auditing score-update dynamics, evaluating measurement instruments for review text and rebuttal interaction, and benchmarking score-increase prediction under temporal and paper-clustered splits. Out-of-scope uses include profiling identifiable participants, deanonymization, live conference decision-making, causal claims about manipulating scores, and training systems to game rebuttals.

The code is in the GitHub repository. All random sampling used seed 50. We verified the release with Python 3.14.3 (\textit{google-generativeai 0.8.6}, \textit{matplotlib 3.10.7}, \textit{networkx 3.5}, \textit{numpy 2.4.3}, \textit{openai 2.29.0}, \textit{openreview-py 1.59.0}, \textit{pandas 3.0.1}, \textit{polars 1.34.0}, \textit{pyarrow 23.0.1}, \textit{python-dotenv 1.2.2}, \textit{scikit-learn 1.8.0}, \textit{scipy 1.17.1}, and \textit{statsmodels 0.14.6}).

\paragraph{Compute resources.}
Rebuilding the released tables and figures from the bundled data is CPU-only. No GPU training was performed; the supervised models are \textit{scikit-learn} random forests/logistic regressions and \textit{statsmodels} logistic regressions. The staged data/results archive is about 5.2 GB uncompressed and 1.3 GB compressed; the consolidated public LLM batch-output directory is about 484 MB. The original score-decoder stage used Gemini Batch API jobs in 37 ICLR 2024--2025 chunks (25 for 2024, 12 for 2025), plus 13 ICLR 2023 chunks used for release coverage, and produced 18,781 paper-level responses parsed into \num{73000} ICLR 2024--2025 review rows. Gemini feature classification used batch outputs for the 1,000-interaction 2025 validation slice and the disjoint 2024 held-out corpus; Claude Opus 4.6 with extended thinking enabled feature induction used five runs over the 1,000-thread induction sample. Provider-side wall-clock time, token accounting, and billing records were not preserved in the public release. Those API calls are not required to rebuild the paper tables and figures because the consolidated model outputs are redistributed.

\section{Dataset Construction and Release Contents}\label{app:dataset-release}\label{app:methods}

The appendix begins with the released measurement resource: corpus construction, review funnel, normalisation, annotation layers, and external before/after score reconciliation used by the headline analyses.

\subsection{Dataset}\label{app:dataset}

\begin{table}[!t]
  \centering
  \caption{\textbf{Author-reply engagement marks a strong selection boundary.} Descriptive comparison among ICLR 2024--2025 reviews with externally matched before/after scores. Author reply means at least one direct author response to the review-rooted thread. Score increase is $\code{ext\_rating\_after} > \code{ext\_rating\_before}$. Deviation is the reviewer's initial score minus the paper-level initial mean.}
  \label{tab:engagement-contrast}
  \small
  \setlength{\tabcolsep}{2pt}
  \begin{tabular}{L{3.4cm}R{1.7cm}R{1.8cm}R{1.8cm}R{2.0cm}R{2.3cm}}
    \toprule
    Author reply status & Reviews & Increases & Increase rate & Mean initial score & Mean deviation from paper mean \\
    \midrule
    No author reply & \num{17990} & 491 & 2.7\% & 4.0 & $-0.32$ \\
    At least one author reply & \num{55241} & \num{14903} & 27.0\% & 5.1 & $+0.10$ \\
    \bottomrule
  \end{tabular}
\end{table}

We collected all publicly visible submissions, reviews, decisions, and discussion data from four years of ICLR (2023--2026) via the OpenReview API. The full corpus contains \num{42049} reviewed papers and \num{164974} raw review records spanning 2023--2026, with \num{436850} text-bearing discussion messages. The main score-update outcome comes from the externally archived before/after comparison documented in \appsec{app:ext-gt}: $\code{score\_increase}=\mathbf{1}[\code{ext\_rating\_after} > \code{ext\_rating\_before}]$, with unchanged and decreased scores both assigned to the negative class. The released measurement layers (OpenReview raw records, external before/after score match, Gemini score-prediction rows, reviewer--author exchange feature labels, and a no-rebuttal subset for engagement contrast) are documented in the data release accompanying this paper.
\paragraph{ICLR 2023: missing withdrawn and desk-rejected papers.}
ICLR 2023 was accessed through OpenReview API v1, which no longer exposes withdrawn or desk-rejected papers. Our 2023 subset therefore contains \num{3793} reviewed papers against \num{4955} total submissions reported by PaperCopilot~\cite{papercopilot}. For 2024 to 2026, our paper counts agree with PaperCopilot to within roughly $1\%$.

\paragraph{ICLR 2026: process-exception year.}
ICLR 2026 is retained in the released longitudinal resource but excluded from score-update analyses. Official ICLR incident reports state that the reviewer discussion was frozen, area chairs were reassigned, and review text and scores were reverted to the beginning of the discussion period after a security incident during rebuttal. A later retrospective states that all review scores were reset to the pre-rebuttal state because the vulnerability began around the rebuttal phase~\cite{iclr2026incident,iclr2026retrospective}. The 2026 rating scale also changed from $\{1,3,5,6,8,10\}$ to $\{0,2,4,6,8,10\}$, and the Spotlight acceptance tier is absent from our 2026 snapshot. The 2026 data should therefore be used as a public-record corpus for that unusual year, not as evidence about ordinary rebuttal-period score updating.

\subsection{Normalization pipeline}\label{app:normalization}

The OpenReview platform underwent structural changes between 2023 (API v1) and 2024--2026 (API v2), and ICLR also modified its review form across years (e.g., \code{recommendation} renamed to \code{rating}, \code{summary\_of\_the\_paper} renamed to \code{summary}, combined \code{strength\_and\_weaknesses} split into separate fields). We implemented a normalisation pipeline using the template method pattern, with a \code{BaseConverter} abstract class and per-conference subclasses handling year-specific field mappings. Participant classification assigns each discussion comment to a type (author, reviewer, area chair, program chair) based on signature patterns. Output files are written as JSON with a manifest recording conversion metadata. The normalised tree contains \num{164974} raw \code{Official\_Review} records and \num{436850} text-bearing discussion comments.

\subsection{External ground truth reconciliation}\label{app:ext-gt}

We use the reviewer-level before- and after-rebuttal scores published by Kargaran et al.~\cite{kargaran} as ground truth for validating both score predictions and score-change detection. \apptab{tab:external-gt} documents the reconciliation between their dataset and ours.

All primary score-update labels in the paper are derived from this external before/after score archive. LLMs are not used to assign the primary score-change outcome.

\begin{table}[!t]
  \centering
  \caption{Reconciliation between our OpenReview dataset and the external ground truth from Kargaran et al. ``With scores'' indicates papers where reviewer-level before/after ratings are available. ``Empty'' papers appear in the external dataset with a title but no reviewer-level data.}
  \label{tab:external-gt}
  \small
  \begin{tabular}{lrrrr}
    \toprule
    & \multicolumn{2}{c}{\textbf{ICLR 2024}} & \multicolumn{2}{c}{\textbf{ICLR 2025}} \\
    \cmidrule(lr){2-3} \cmidrule(lr){4-5}
    & Papers & Reviews & Papers & Reviews \\
    \midrule
    Our dataset (all)              & \num{7404}  & \num{28028}  & \num{11672} & \num{46748}  \\
    \quad with reviews             & \num{7262}  & \num{28028}  & \num{11520} & \num{46748}  \\
    External GT (total entries)    & \num{7407}  &              & \num{11677} &              \\
    \quad with reviewer scores     & \num{6973}  & \num{26878}  & \num{11431} & \num{46353}  \\
    \quad empty (no reviewer data) & 434         & --           & 246         & --           \\
    \midrule
    Matched (reviews in both)      &             & \num{26878}  &             & \num{46353}  \\
    Only in our data               &             & \num{1150}   &             & 395          \\
    \bottomrule
  \end{tabular}
\end{table}

\paragraph{Papers without reviewer scores in the external dataset.}
The external dataset contains entries for \num{7407} ICLR 2024 papers and \num{11677} ICLR 2025 papers. However, 434 (2024) and 246 (2025) of these entries have no reviewer-level score data: the paper title and tracing score are present, but the review dictionary is empty. These 680 empty entries fall into two distinct categories.

\textit{Case 1: papers with reviews that the external scraper missed (379 papers).} Of the 680 empty entries, 289 in 2024 and 90 in 2025 correspond to papers that do have reviews in our dataset and on the OpenReview website. These 379 papers account for \num{1545} reviews for which we have the final score and Gemini predictions but no before/after ground truth. We verified a sample by re-querying the OpenReview API (April 2026): all sampled papers have full review data publicly accessible, so the missing entries reflect coverage of the external before/after scrape rather than the underlying OpenReview records. These papers do not differ systematically from covered papers on decision distribution (54.8\% Rejected vs.\ 47.3\%), mean rating (5.15 vs.\ 5.14), review dates, or topic areas.

\textit{Case 2: withdrawn or desk-rejected papers with no reviews anywhere (301 papers).} The remaining 145 (2024) and 156 (2025) empty entries correspond to papers that also have zero reviews in our data and on the OpenReview website. Neither dataset has review data for them because no reviews were ever written.

\paragraph{OpenReview ratings reflect post-rebuttal scores.}
The rating returned by the OpenReview API reflects the reviewer's final post-rebuttal score. It does not record the initial score. We confirmed this by comparing the OpenReview rating against the after-rebuttal score from the external ground truth: 100\% exact match across all \num{26878} matched reviews in ICLR 2024 and all \num{46353} matched reviews in ICLR 2025. This means that comparing LLM score predictions against OpenReview ratings would conflate two effects: prediction error and rebuttal-induced score changes. We therefore compare the decoded score from score-stripped archived review text to the external pre-rebuttal score. The strict pre-response interpretation is supported by H0-minimal, which excludes mutable review-text-derived features. All prediction-accuracy figures in this paper use the initial pre-rebuttal score as ground truth.

\paragraph{Review-count discrepancies.}
For ICLR 2024, our dataset contains \num{28028} reviews while the external dataset contains \num{26878} reviewer-level records. The difference of \num{1150} reviews comes entirely from the 289 papers with empty review records in the external data. Within papers that both datasets cover, the reviewer-level match is exact. For ICLR 2025, the reviewer-level match is also exact: all \num{46353} reviewer records in the external dataset are matched in our data.

\section{Review-Text Measurement}\label{app:review-text-measurement}

The review-text appendix covers the score decoder, the text--score offset, auxiliary score-change mention detection, and inter-reviewer agreement context.

\subsection{Gemini score-prediction protocol}\label{app:gemini-pred}

\paragraph{Task.}
We tested whether a Gemini Flash 3.0 model can infer reviewer scores from archived review text alone, without access to any numeric rating fields. We submitted all ICLR 2024 and 2025 archived reviews to the Gemini Batch API with a structured prompt requesting per-reviewer inferred scores on the ICLR discrete scale $\{1, 3, 5, 6, 8, 10\}$, along with confidence, sentiment, and engagement-quality ratings. The analysis treats the resulting outputs as one Gemini Flash 3.0 score-decoder instrument rather than as a model-comparison experiment. Temperature was set to 0.0 for deterministic output.

\paragraph{Score-field stripping.}
Before sending reviews to the LLM, we removed all numeric scoring fields to prevent the model from reading the score rather than inferring it from the review text. The strip set covers, across all years: \code{soundness}, \code{presentation}, \code{contribution}, \code{rating}, \code{recommendation}, and \code{confidence} (the reviewer's self-reported confidence on a 1--5 scale). For ICLR 2023 reviews, the strip set additionally covers the year-specific sub-scores: \code{correctness}, \code{technical\_novelty\_and\_significance}, \code{empirical\_novelty\_and\_significance}, and \code{clarity,\_quality,\_novelty\_and\_reproducibility}. Matching is case- and space-insensitive (e.g., ``Confidence'' and \code{confidence} both match) to handle formatting variations across years. After stripping, the fields remaining in the input dict for an ICLR 2024 to 2025 review are the four free-text fields written by the reviewer (\code{summary}, \code{strengths}, \code{weaknesses}, \code{questions}) plus the boolean ethics flag (\code{flag\_for\_ethics\_review}). The boolean flag is binary metadata that cannot leak the score.

\paragraph{Batch processing.}
ICLR 2024 was processed in 25 chunks (February to March 2026), ICLR 2023 in 13 chunks (March 14 to 15, 2026), and ICLR 2025 in 12 chunks (March 24, 2026). Papers without reviews were excluded: 142 in ICLR 2024, 152 in ICLR 2025. Project-level provenance records list Gemini Flash 3.0 for the ICLR 2023, 2024, and 2025, all at temperature 0.0. Neither the batch input files nor the row-level output files contain model metadata, so the slim release treats exact Gemini model strings as project-level provenance rather than row-level covariates.

\paragraph{Validation against the initial pre-rebuttal score.}
We validated predictions against the initial pre-rebuttal scores from the external ground truth (\appsec{app:ext-gt}), before the final post-rebuttal scores became visible on OpenReview. Combined across ICLR 2024 and 2025 ($n = \num{73000}$), exact match is 40.7\%, within-one-step is 87.6\%, and Spearman $\rho = 0.40$. By year, ICLR 2024 reaches 36.9\% / 81.2\% / $\rho = 0.330$ ($n = \num{26805}$), and ICLR 2025 reaches 42.9\% / 91.2\% / $\rho = 0.457$ ($n = \num{46195}$). We report Spearman $\rho$ rather than Pearson $r$ because the ICLR scale has uneven step sizes (\{1, 3, 5, 6, 8, 10\} with gaps 2, 2, 1, 2, 2). Pearson's magnitude depends on the numeric spacing of the scale, while Spearman is rank-based and invariant to it.

\paragraph{Offset operationalisation.}
The text--score offset used throughout \Cref{res:instrument,sec:horizons,sec:rebuttal-impact} and \appsec{app:text-score-gap} is computed as a difference of step positions on the six-point ICLR ladder instead of raw rating points. Concretely, both the Gemini-decoded score and the assigned initial score are first mapped through the position table $\{1{:}0,\, 3{:}1,\, 5{:}2,\, 6{:}3,\, 8{:}4,\, 10{:}5\}$, and the offset is the difference of indices. Two reviewers who are ``one step apart'' on the ladder always have $|\code{offset}| = 1$ regardless of where on the scale they sit, so the $5 \to 6$ adjacent pair is treated identically to the $8 \to 10$ adjacent pair. Reading the offset in raw rating points would conflate the $5{-}6$ jump (one ladder step but one raw point) with the $6{-}8$ jump (also one ladder step but two raw points) and bias the regression coefficients accordingly. The mean offset of $+0.14$ and standard deviation of $1.02$ reported throughout the paper are therefore in step units. The offset bins in \Cref{fig:f2-instrument} collapse the offset to its sign only (\code{under} for $\code{offset} < 0$, \code{match} for $\code{offset} = 0$, \code{over} for $\code{offset} > 0$). The regressions of \Cref{sec:horizons,sec:rebuttal-impact} use the integer step difference standardised by year-pooled mean and standard deviation.

\paragraph{Distributional fidelity is the right yardstick.}
The within-one-step accuracy of 87.6\% is not load-bearing evidence on its own. A constant predictor that always emits the modal score 5 reaches 88.8\% within one step on the same data, simply because the assigned scores cluster around 5--6. That constant predictor cannot produce a per-review prediction that varies with the input text, which is what the analyses in \Cref{res:instrument} require. Two pieces of evidence speak directly to per-review informativeness. First, Gemini's prediction distribution spans the ICLR scale and roughly tracks the assigned distribution: 0.4\% / 19\% / 46\% / 27\% / 7\% / 0.0\% on \{1, 3, 5, 6, 8, 10\} for predictions vs.\ 2.4\% / 30\% / 35\% / 24\% / 9\% / 0.3\% for assigned scores. Second, the LLM-decoded-minus-assigned offset predicts post-rebuttal score-increase rates of 8.3\% / 19.2\% / 31.9\% across the three offset bins (\appsec{app:text-score-gap}), which is information no constant predictor can encode because its offset would be a deterministic function of the assigned score alone.

\subsection{Text-versus-score gap analysis}\label{app:text-score-gap}

\paragraph{Operationalisation.}
The offset is the Gemini-decoded score (from text) minus the actually assigned initial score, both projected onto the six-step ICLR rating index $\{0, 1, 2, 3, 4, 5\}$ corresponding to scores $\{1, 3, 5, 6, 8, 10\}$. We index step positions rather than raw points because the ICLR scale has uneven gaps (\appsec{app:dataset}), so a reviewer who scored 5 (step 2) and was decoded as 6 (step 3) has offset $+1$, regardless of the unequal raw-point distance. Reviews with either side outside the six-point set are dropped. Reviews are pooled across ICLR 2024 and 2025, with $n = \num{73000}$ where the external ground truth is available.

\paragraph{Offset gradient.}
The 73,000 reviews split into 24.3\% under (Gemini below reviewer), 40.7\% match, and 35.0\% over (Gemini above reviewer). The full three-outcome breakdown (\apptab{tab:offset-gradient}) shows the increase gradient and a mirror decrease gradient running in the opposite direction. The direction of both gradients replicates in 2024 and 2025. The magnitude of the increase gradient is larger in 2025 (8.5\% to 35.8\%) than in 2024 (7.9\% to 26.1\%).

\begin{table}[!t]
    \centering
    \caption{Post-rebuttal outcome distribution by offset bin (combined ICLR 2024 + 2025, $n = \num{73000}$). The increase gradient (8.3 to 31.9\%) and decrease gradient (3.3 down to 0.4\%) run in opposite directions, consistent with the offset capturing bidirectional volatility rather than just an increase tendency.}
    \label{tab:offset-gradient}
    \begin{tabular}{lrrrr}
    \toprule
    Offset bin & $n$ & Increase & Same & Decrease \\
    \midrule
    Under (Gemini $<$ assigned) & \num{17736} & 8.3\% & 88.4\% & 3.3\% \\
    Match (equal)               & \num{29706} & 19.2\% & 79.8\% & 1.0\% \\
    Over  (Gemini $>$ assigned) & \num{25558} & 31.9\% & 67.6\% & 0.4\% \\
    \midrule
    All                         & \num{73000} & 21.0\% & 77.6\% & 1.4\% \\
    \bottomrule
    \end{tabular}
\end{table}

\paragraph{Covariate-adjustment robustness check.}
Adding Gemini's per-review engagement, sentiment, and tone ratings jointly as covariates to the logistic increase model shrinks the offset coefficient by 23.5\% ($c = 0.643$ to $c' = 0.492$, $n = \num{72990}$). About 76\% of the offset's association with later increase is therefore unexplained by these auxiliary ratings. We report this as a covariate-adjustment robustness check and make no formal mediation claim.

\section{Resolved-Thread Feature Taxonomy}\label{app:thread-taxonomy}

The taxonomy appendix gives the induction sample, validation protocol, multiple-testing checks, and conservative core for the 44 resolved-thread features.

\subsection{Resolved-thread feature coding: 44-feature taxonomy}\label{app:rebuttal-coding}

The feature taxonomy was induced and validated in three stages: induction with LLM-A (Claude Opus 4.6 with extended thinking enabled), independent classification with LLM-B (Gemini Flash 3.0), and a held-out replication test on disjoint data. Two conditions were declared at the start of the analysis: (a) the features reflect structure in the interaction text rather than patterns internal to LLM-A, and (b) the feature-outcome associations replicate beyond the induction sample.

\paragraph{Sample for taxonomy induction.}
From ICLR 2025 we drew \num{1000} reviewer-paper pairs stratified by reviewer initial rating (low 1--3, medium 5--6, high 8--10) crossed with rebuttal outcome (\code{increase} vs.\ \code{same\_decrease}), restricted to pairs with a non-empty author rebuttal thread. Each pair was serialised as a raw text thread containing the full original review (summary, strengths, weaknesses, questions), the full author rebuttal, and any subsequent reviewer and author follow-ups, in chronological order. No structured Gemini-extracted fields (sentiment, tone, engagement-quality, score-change verdict outputs) were exposed to either LLM-A or LLM-B at any point in this analysis. The serialised threads are included in the public release. The high-increase cell is rare in ICLR 2025. We therefore used all 58 available pairs rather than subsample, which matches the strategy later applied to the ICLR 2024 hold-out. The remaining five cells were filled by random sampling.

\paragraph{Feature induction (LLM-A = Claude).}
In a single prompt we presented LLM-A with all six cells of the sample and instructed the model to perform the contrast within each tier (low-\code{increase} vs.\ low-\code{same\_decrease}, and analogously for medium and high). The within-tier framing preserves tier-specific patterns and prevents the model from collapsing the task into a single cross-tier contrast that could be dominated by score-level differences rather than resolved thread content. We repeated the procedure five times to attenuate run-to-run variability, then manually aggregated the extracted categories across runs and tiers, merging entries that were pure rephrasings or semantically close. The aggregation yielded 44 features, each labelled with a declared polarity (positive or negative) from the pooled induction output.

\paragraph{Memory-control swaps.}
To rule out the hypothesis that LLM-A was retrieving canonical patterns from its prior rather than reading the interactions, we re-ran the induction under two manipulations. The first swapped the low-tier and high-tier interactions within the same prompt structure. The second silently swapped the \code{increase} and \code{same\_decrease} outcome labels on the same \num{1000} interactions, holding tier assignment fixed. In both cases the resulting feature content shifted with the presented inputs rather than tracking the labels, consistent with extraction from the interaction text. The outcome-label swap is the sharper of the two stress tests, since it operates on the axis the original induction used to discriminate. It confirms that LLM-A continues to locate the same features on the same actual data even when the labels are flipped, with only the surface narrative re-framing around the new labels.

\paragraph{Classification protocol (LLM-B = Gemini Flash 3.0).}
A second, independent Gemini call classified each interaction against the fixed 44-feature list. The classifier saw the same raw thread text used at induction (no structured features), the 44-feature taxonomy with shuffled per-request feature identifiers, and the request to mark which features were present. Classification was outcome-blinded: reviewer tier and rebuttal outcome were stripped from the prompt, and feature identifiers were shuffled per request so that ordering could not cue the model. Because the batch JSONL endpoint did not accept Gemini structured-output fields for these requests, the closed set was enforced by prompt instructions and downstream validation rather than by a response schema: requests instructed the model to use only F01--F44, and parsing discarded malformed JSON or feature IDs outside that set. For every feature we computed an odds ratio comparing the \code{increase} and \code{same\_decrease} outcome groups, with a 95\% confidence interval, and a two-sided permutation test with 2000 permutations. Under each permutation the \code{increase} / \code{same\_decrease} labels are randomly reshuffled across interactions to simulate the null of no feature-outcome association, the odds ratio is recomputed, and the $p$-value is the fraction of shuffles whose $|\log \text{OR}|$ matches or exceeds the observed value (with add-one smoothing to keep $p > 0$). Significance is reported at the Bonferroni-corrected threshold $\alpha = 0.05 / 44 = 0.0011$. Choosing 2000 permutations gives a resolution floor of about $1/2001 \approx 0.0005$, just below the Bonferroni-corrected threshold.

\paragraph{Choice of multiple-testing correction.}
Bonferroni is the most stringent of the standard family-wise error-rate (FWER) controls and assumes nothing about test independence. We anchor on it because the held-out replication on a disjoint ICLR 2024 sample (Test 2) is itself a strong robustness check, and because surviving the strictest correction leaves least room for over-claiming. Two alternative corrections are reasonable, both more permissive (\apptab{tab:mtcorrection}). Holm-Bonferroni controls the same FWER as Bonferroni but is uniformly more powerful~\cite{holm1979simple}, and it yields 25 doubly-significant features instead of 23. The Benjamini-Hochberg FDR rule controls the expected proportion of false positives among the rejected hypotheses, a weaker but often more appropriate guarantee for discovery-style analyses~\cite{benjamini1995controlling}. At $q = 0.05$, it yields 31 doubly-significant features. Direction of effect agrees with the Bonferroni-anchored core under all three corrections, and the more permissive corrections add features to the core rather than displace any of its members.

\begin{table}[!t]
    \centering
    \caption{Sensitivity of the doubly-significant feature count to the multiple-testing correction. Bonferroni is the headline reported in \Cref{res:taxonomy}.}
    \label{tab:mtcorrection}
    \begin{tabular}{lccc}
    \toprule
    Correction & Test 1 & Test 2 & In both \\
    \midrule
    Uncorrected $p < 0.05$              & 32/44 & 43/44 & --- \\
    Bonferroni $\alpha = 0.05 / 44$    & 23/44 & 40/44 & \textbf{23/44} \\
    Holm-Bonferroni $\alpha = 0.05$    & 25/44 & 42/44 & 25/44 \\
    Benjamini-Hochberg $q = 0.05$      & 31/44 & 43/44 & 31/44 \\
    Benjamini-Hochberg $q = 0.10$      & 32/44 & 44/44 & 32/44 \\
    \bottomrule
    \end{tabular}
\end{table}

\paragraph{Tests.}
\textbf{Test 1 (operationalisability)} applies LLM-B to the same \num{1000} ICLR 2025 interactions used for induction, asking whether a second model can recover the induced features from the same text. \textbf{Test 2 (discriminative validity)} applies the same protocol to the full classified ICLR 2024 cohort of \num{5730} interactions disjoint from the induction corpus, asking whether the feature-outcome associations replicate on fresh data. An earlier round of Test 2 was reported on a \num{952}-interaction stratified subsample of the same cohort (six-cell stratification by tier and verdict-group). The conclusions of that smaller test (43 of 44 declared polarities preserved, 16 of 44 doubly-Bonferroni-significant) hold under the present full-cohort version, with strict strengthening on every count.

\paragraph{Cohort thinness and clustering.}
The classified ICLR 2024 cohort contains \num{5730} rebuttals across \num{3460} unique papers, an average of $1.66$ rebuttals per paper. This is a thinned slice of ICLR 2024's natural ${\sim}4$-reviewer paper structure, because not every reviewer wrote a rebuttal and not every (paper, reviewer) pair survived Gemini classification. The headline Test 2 permutation test uses naive (row-level) shuffling, matching the protocol used for the original Test 1 / Test 2 pair. As a robustness check we re-ran the expanded Test 2 with a within-paper block permutation that swaps outcome blocks at the paper level. The doubly-Bonferroni-significant count is unchanged at $23$ and per-feature $p$-values agree to within $0.005$. We do not lean on the block-permutation result as a paper-clustered correction in the body text, because the cohort thinness limits how much within-paper correlation our block test can detect, and we cannot rule out residual within-paper correlation in unobserved rebuttals or selection bias in which (paper, reviewer) pairs produced a classified rebuttal.

\paragraph{Results.}
The taxonomy shows the declared directional structure in both tests. In Test 1, 43 of 44 features are more prevalent on the side of the \code{increase} versus \code{same\_decrease} contrast declared at induction (22 of 23 positive, 21 of 21 negative), and 23 of 44 survive the Bonferroni threshold. In Test 2, all 44 of 44 features match their declared direction (23 of 23 positive, 21 of 21 negative), and 40 of 44 individually survive Bonferroni at the full-cohort sample size. The single Test 1 disagreement is F12 (proactive limitation acknowledgement), which inverts polarity with OR = 0.90 [0.68, 1.19]. Under the original \num{952}-interaction Test 2 subsample, F40 (marginal quantitative improvements) was the corresponding inversion at OR = 1.52 [0.76, 3.05], but it lands at OR = 0.36 [0.26, 0.51] in the full cohort, matching its declared negative polarity. Of the 44 features, 23 cross Bonferroni in both tests (9 positive, 14 negative) and form the conservative core listed in \apptab{tab:taxonomy-core}. No feature crosses only in Test 1, and a further 17 cross only in Test 2 (the larger sample resolves several borderline features that the in-sample run lacks power to detect). These are retained but reported as weaker evidence.

\begin{table}[!t]
  \centering
  \caption{The 23 doubly-significant features that anchor the taxonomy. ``Polarity'' is the direction declared at induction (LLM-A) and confirmed by both LLM-B tests at the Bonferroni-corrected threshold $\alpha = 0.05 / 44$.}
  \label{tab:taxonomy-core}
  \small
  \begin{tabular}{p{0.8cm} p{1.5cm} p{11cm}}
    \toprule
    \textbf{ID} & \textbf{Polarity} & \textbf{Label} \\
    \midrule
    F02 & positive & Non-defensive, respectful, collaborative tone. \\
    F04 & positive & Sincere engagement: detailed responses, length, added appendix content. \\
    F05 & positive & Methodological detail and reproducibility clarification. \\
    F06 & positive & Concrete empirical evidence tied to reviewer concerns. \\
    F07 & positive & Baseline completion / comparator strengthening. \\
    F13 & positive & Explicit manuscript revision signaling. \\
    F14 & positive & Supplementary material and new visual evidence. \\
    F21 & positive & Active dialogue and continued engagement. \\
    F22 & positive & Concrete revision commitments. \\
    \midrule
    F24 & negative & Defensive or dismissive tone. \\
    F25 & negative & Superficial or partial addressal of reviewer concerns. \\
    F27 & negative & Missing or insufficient baseline comparisons. \\
    F28 & negative & Lack of quantitative / empirical evidence in rebuttal. \\
    F29 & negative & No new requested experiments / analyses / ablations. \\
    F30 & negative & Vague future-oriented promises instead of concrete commitments. \\
    F31 & negative & Narrow or insufficient evaluation scope. \\
    F32 & negative & Incomplete empirical coverage even after rebuttal additions. \\
    F33 & negative & Unresolved quality-of-execution concerns that further experiments do not fix. \\
    F36 & negative & Weak novelty justification. \\
    F38 & negative & Section-referencing or paper-rephrasing instead of real clarification. \\
    F39 & negative & Inadequate explanation or persistent clarity issues. \\
    F42 & negative & Vague or insufficient technical explanation. \\
    F43 & negative & No explicit commitment to revise or improve the paper. \\
    \bottomrule
  \end{tabular}
\end{table}

\begin{figure}[!p]
  \centering
  \includegraphics[width=\textwidth,height=0.82\textheight,keepaspectratio]{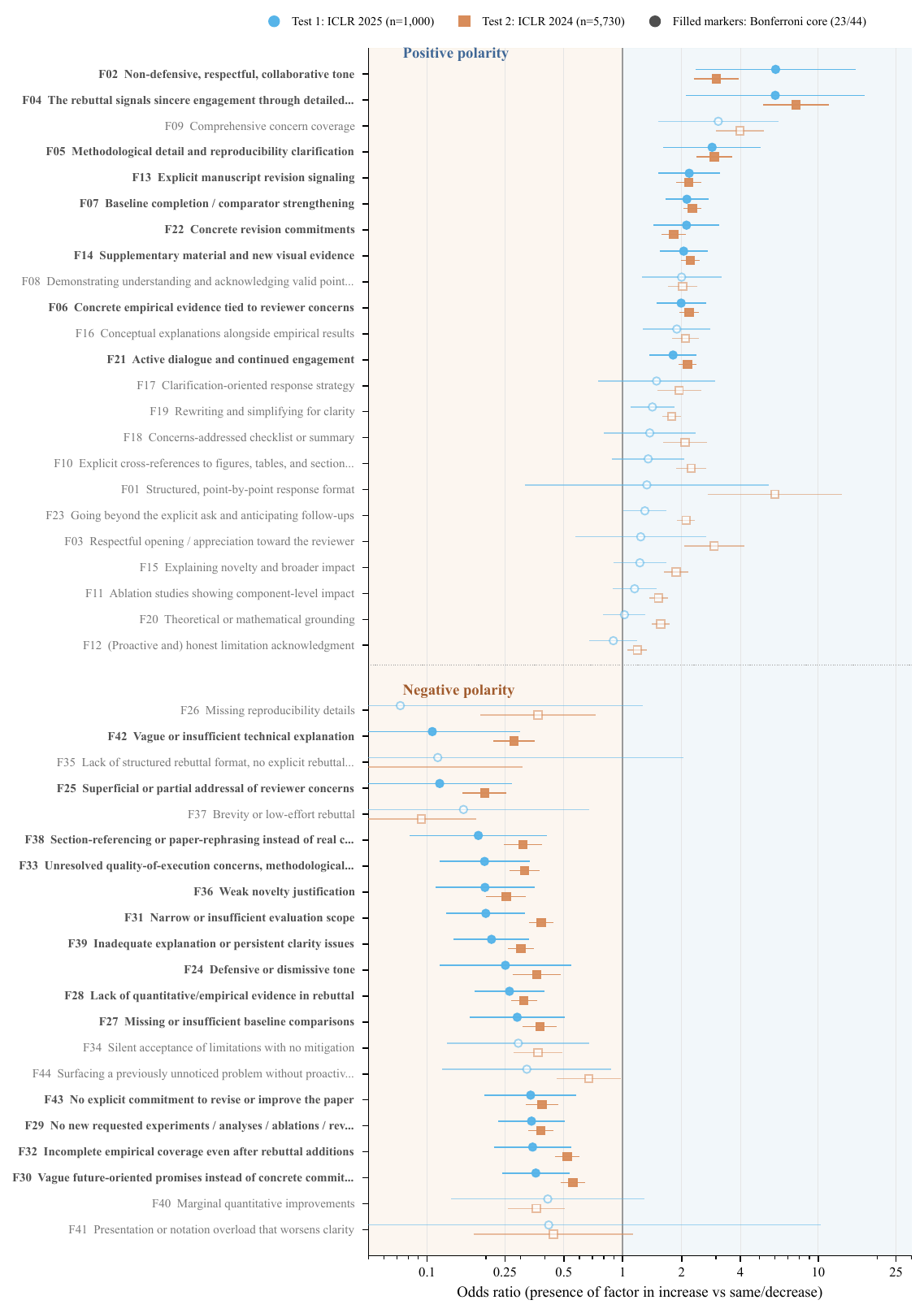}
  \caption{\textbf{Full 44-feature exchange-codebook validation forest.} Each row is one induced feature, plotted as an odds ratio for score increase versus same/decrease (log scale). Test 1: ICLR 2025 induction sample ($n=\num{1000}$, blue circles). Test 2: held-out ICLR 2024 interactions ($n=\num{5730}$, orange squares). Filled markers mark the 23 features shown in the main-text Bonferroni-core forest (\Cref{fig:f3-categories}).}
  \label{fig:f3-categories-full-app}
\end{figure}

\section{Prediction Horizons and Conditional Inference}\label{app:prediction-inference}

The prediction appendix defines the three observation horizons: initial review (H0), resolved thread (H1), and panel discussion (H2). It then reports the conditional feature model, robustness checks, calibration, and model stress tests.

\subsection{Three-horizon prediction and inferential modelling}\label{app:horizons}

\paragraph{Outcome.} $\code{score\_increase} = \mathbf{1}[\code{ext\_rating\_after} > \code{ext\_rating\_before}]$. \code{Same} and \code{decrease} collapse to the negative class. Base rate on the temporal-test set is $44.4\%$. The 2025 slice is stratified and rare-cell-enriched across tier and verdict group. On the year-2024 training set the base rate is $45.8\%$ on the same denominator.

\paragraph{Cohort.} All horizons share the same row denominator: every reviewer-paper pair where (i) the parquet records both $\code{ext\_rating\_before}$ and $\code{ext\_rating\_after}$, (ii) Gemini produced a per-review prediction, (iii) the rebuttal-features extractor records $\code{has\_rebuttal} = 1$, and (iv) the $F_{01}\ldots F_{44}$ classifier produced a parsed label set. We use two related ICLR 2024 counts. The feature-validation Test 2 cohort of \Cref{res:taxonomy} and \appsec{app:rebuttal-coding} is $n = \num{5730}$ rebuttals where Gemini produced a parsed feature label set (the disjoint validation set on which per-feature odds ratios are reported). The three-horizon prediction cohort used here is $n = \num{5710}$, which is the $\num{5730}$ feature-classified rebuttals after intersecting with the additional benchmark requirements (Gemini score-prediction, $\code{has\_rebuttal}$, and both external scores). The 2025 cohort is the stratified, rare-cell-enriched $n=995$ Test 1 sample. A budgeted follow-up to extend the $2025$ cohort beyond the stratified slice failed at submission time and was not re-attempted within the project budget. The temporal-split AUCs in \Cref{sec:rebuttal-impact} are therefore measured on a stratified validation slice rather than a population-representative 2025 draw. The $\texttt{GroupKFold}$ AUCs use both years and are not affected.

\begin{table}[!t]
  \centering
  \caption{\textbf{H0 features differ in provenance.} Provenance audit for the initial-review and thread-level feature families. H0-minimal is the conservative ablation used to test the headline initial-review claim without mutable OpenReview review text or LLM-derived review-text measurements.}
  \label{tab:h0-provenance}
  \small
  \setlength{\tabcolsep}{3pt}
  \begin{tabular}{L{3.5cm}L{4.6cm}L{3.2cm}L{3.4cm}}
    \toprule
    Feature family & Variables & Provenance & Use in leakage-sensitive interpretation \\
    \midrule
    Outcome & $\code{ext\_rating\_after} > \code{ext\_rating\_before}$ & External before/after score archive & Primary label only, never an LLM output \\
    Initial score context & $\code{pos\_before}$, paper initial mean/std, reviewer deviation, reviewer count & External pre-rebuttal score archive & Included in H0-minimal and full H0 \\
    Review timing & $\code{review\_cdate}$, $\code{days\_before\_deadline}$ & OpenReview creation metadata & Included in H0-minimal and full H0 \\
    Review-form fields & confidence, soundness, presentation, contribution, review-text lengths & Downloaded OpenReview note content & Included in full H0, excluded from H0-minimal because released records do not prove initial-revision text for every field \\
    Review-text measurements & Gemini text--score offset, tone, engagement indicators & Gemini over score-stripped downloaded review text & Included in full H0, excluded from H0-minimal because they inherit the review-text provenance caveat \\
    Resolved thread & rebuttal aggregates, thread depth, feature count, $F_{01}\ldots F_{44}$ & Parsed review-rooted thread including author replies and reviewer follow-ups & H1 resolved-thread measurement, not author-first-response measurement \\
    Later discussion & total thread messages and co-reviewer movement & Post-discussion process observables & H2 retrospective benchmark only \\
    \bottomrule
  \end{tabular}
\end{table}

\paragraph{Feature buckets.} The prediction benchmark uses three nested feature sets and one leakage-sensitive H0 ablation. H0-minimal contains $\code{pos\_before}$, $\code{paper\_initial\_mean}$, $\code{paper\_initial\_std}$, $\code{n\_reviewers}$, $\code{deviation\_from\_paper\_mean}$, and $\code{days\_before\_deadline}$. The full initial-review horizon (H0, 30 features after one-hot expansion) contains those variables plus $\code{confidence}$, $\code{soundness}$, $\code{presentation}$, $\code{contribution}$, $\code{strengths\_len}$, $\code{weaknesses\_len}$, $\code{questions\_len}$, $\code{sw\_ratio}$, $\code{is\_initial\_outlier}$ ($|\code{deviation}| \ge 1$ step and $\code{confidence} \ge 4$), $\code{gemini\_offset}$, $\code{engagement\_ord}$, $\code{score\_confidence\_gemini}$, $\code{n\_indicators}$, four binary engagement indicators, and the one-hot expansion of $\code{tone\_primary}$. The Gemini confidence scalar is retained only as part of the pre-specified Gemini metadata block and is not interpreted as a validated standalone channel. The resolved-thread horizon (H1, 79 features) adds $\code{rebuttal\_total\_len}$, $\code{n\_author\_replies}$, $\code{first\_reply\_latency\_days}$, $\code{thread\_max\_depth}$, $\code{n\_features\_present}$, and the 44 binary $F_{01}\ldots F_{44}$ indicators. The panel-discussion horizon (H2, 83 features) adds $\code{n\_thread\_messages}$ plus $\code{co\_reviewer\_post\_change\_mean}$, $\code{co\_reviewer\_max\_increase}$, and $\code{co\_reviewer\_n\_increased}$, all computed over the other reviewers of the same paper.

\paragraph{Leakage audit.} We ran four leakage checks:
\begin{enumerate}[itemsep=0pt,topsep=2pt]
  \item the H0 feature set contains none of the column names from the H1- or H2-only buckets.
  \item the H0-minimal matrix contains only the six provenance-minimal features listed in \apptab{tab:h0-provenance}.
  \item synthetic perturbation: changing any $\code{ext\_rating\_after}$ value leaves $\code{paper\_initial\_mean}$ unchanged, ruling out accidental use of post-rebuttal scores in the paper-level summary.
  \item no individual feature has $\lvert r \rvert \ge 0.99$ with the binary outcome at any horizon (outcome-leak tripwire).
\end{enumerate}
All four tests pass at the time of writing.

\paragraph{Models.} Logistic regression: $\ell_2$-regularised, $\code{class\_weight="balanced"}$, $\code{max\_iter}=2000$, features standardised at fit time. Random forest: 500 trees, $\code{min\_samples\_leaf}=20$, $\code{class\_weight="balanced\_subsample"}$, $\code{n\_jobs}=-1$. Both fitted at all three horizons.

\paragraph{Evaluation.} (i) Temporal split: train on the year-$2024$ cohort, test on the year-$2025$ stratified slice. (ii) $\texttt{GroupKFold}(5)$ on the combined $n=6{,}705$ cohort, with $\code{paper\_key}$ year-prefixed so a $\code{paper\_key}$ shared between years cannot leak across folds. Both metrics ($\auc$ and $\auprc$) are reported. Non-parametric bootstrap with $1000$ resamples gives the temporal-split CIs. Permutation importance ($n_{\text{repeats}}=10$, scoring $=$ average precision) is computed on the temporal-test set per horizon.

\begin{table}[!t]
  \centering
  \small
  \caption{Predictive performance across observation horizons. Temporal split trains on ICLR 2024 and tests on the stratified, rare-cell-enriched ICLR 2025 slice. GroupKFold uses paper-clustered folds on the combined $n=6{,}705$ cohort. Values correspond to the feature buckets summarised in \Cref{fig:content-lift}.}
  \label{tab:horizon-metrics}
  \setlength{\tabcolsep}{4pt}
  \begin{tabular}{llcccc}
    \toprule
    Horizon & Model & Temporal AUC & Temporal AUPRC & CV AUC & CV AUPRC \\
    \midrule
    H0 initial-review        & Logistic      & 0.747 & 0.724 & 0.779 & 0.746 \\
    H0 initial-review        & Random forest & 0.724 & 0.705 & 0.781 & 0.737 \\
    \addlinespace
    H1 resolved thread & Logistic      & 0.804 & 0.780 & 0.848 & 0.818 \\
    H1 resolved thread & Random forest & 0.769 & 0.752 & 0.835 & 0.800 \\
    \addlinespace
    H2 panel discussion      & Logistic      & 0.809 & 0.784 & 0.857 & 0.832 \\
    H2 panel discussion      & Random forest & 0.771 & 0.744 & 0.842 & 0.811 \\
    \bottomrule
  \end{tabular}
\end{table}

\paragraph{Inferential logistic with cluster-robust SE.} The headline model behind \Cref{sec:horizons} and \Cref{fig:recforest} is fit with \code{statsmodels.Logit} using $\code{cov\_type="cluster"}$ grouped by $\code{paper\_key}$, on the combined 2024+2025 resolved thread cohort ($n=6{,}705$, base rate $45.6\%$). Continuous features are standardised by year-pooled mean and standard deviation. Binary features ($F_{01}\ldots F_{44}$, the indicators, and tone one-hots) are kept on $\{0, 1\}$. We drop one tone column as the reference category (the most prevalent in the cohort, $\code{tone\_enthusiastic}$), and drop binary features with prevalence $<1\%$ or $>99\%$ (this removes $F_{01}$ at $99.0\%$ prevalence and three rare negatives $F_{26}$, $F_{35}$, $F_{41}$ at $<1\%$). The pseudo-$R^2$ of the headline fit is $0.312$. The structural-confounder block includes a timing control, $\code{days\_before\_deadline}$, defined as the official ICLR review-submission deadline minus $\code{review\_cdate}$ (in days, with positive values for on-time reviews and negative values for late submissions). The deadlines are 2023-11-10 23:59 UTC for ICLR 2024 and 2024-11-12 23:59 UTC for ICLR 2025, the dates published on the corresponding OpenReview venue calendars. In the two cohorts, $99.0\%$ and $99.7\%$ of recorded $\code{review\_cdate}$ values fall on or before these times. The remaining $0.5\%$ of cohort reviews are kept with negative deadline distances rather than excluded. This operationalisation matches Jung et al.'s ``days remaining until deadline''~\cite{jung} directly and is also consistent with Kargaran et al.'s ``temporal patterns in review submissions''~\cite{kargaran}. Its conditional contribution to the joint fit is per 1 SD OR $= 1.19$ [$1.04$, $1.36$], $p = 0.011$ (earlier-submitted reviews are more likely to result in a post-rebuttal increase). All variance-inflation features in the fitted joint are below 5 ($\code{rebuttal\_total\_len} = 4.57$, $\code{pos\_before} = 4.30$, $\code{first\_reply\_latency\_days} = 4.06$, $\code{days\_before\_deadline} = 4.06$, $\code{n\_author\_replies} = 4.03$). The moderate latency-vs-deadline overlap (Pearson $r = 0.86$ at the raw-feature level) is absorbed by the 44-feature block in the full design. The resolved thread structural block adds two further Jung-style features: $\code{first\_reply\_latency\_days}$ (cdate of the earliest direct author reply minus the review cdate, in days) and $\code{thread\_max\_depth}$ (longest root-to-leaf path of the reply tree rooted at the review). The first does not clear conventional significance in the joint (OR $=0.90$ [$0.79$, $1.03$], $p = 0.12$). The second carries an independently significant signal (OR $=1.24$ [$1.15$, $1.33$], $p < 10^{-8}$). We exclude two resolved thread aggregates from this inferential joint by design. First, $\code{n\_thread\_messages}$ counts all messages in the review-rooted thread, including reviewer follow-ups, and lives in the H2 panel discussion horizon of \Cref{sec:rebuttal-impact}, where it is added to the predictive feature set together with the co-reviewer-movement aggregates. Second, $\code{n\_features\_present}$ would condition every per-feature OR on a fixed total feature count, which is awkward to interpret as ``presence vs absence''. The two aggregates are treated asymmetrically across the inferential and predictive analyses. $\code{n\_thread\_messages}$ is excluded from the inferential joint of \Cref{sec:horizons} and from the H1 predictive horizon of \Cref{sec:rebuttal-impact}. $\code{n\_features\_present}$ is excluded from the inferential joint only and is retained in the H1 and H2 predictive horizons as one additional resolved thread scalar. The asymmetry is intentional because the inferential analysis prioritises clean per-feature coefficients, while the prediction benchmark prioritises rank-based recovery of the binary outcome and tolerates feature redundancy. The calibration-tested predictive logistic inherits the inferential exclusions, whereas the benchmark summarised in \Cref{fig:content-lift} uses the slightly richer feature set described above. The 44 feature indicators themselves are extracted by Gemini from the full reviewer--author thread (author rebuttal plus any subsequent reviewer and author follow-ups, \Cref{res:taxonomy}). The H1 label is therefore ``resolved thread.'' It marks the temporal scope without treating the indicators as properties of the author's first response alone. We do not currently re-run the classification on author-first-response-only excerpts. Given the cost incurred during the original Gemini batch, this remains a documented limitation rather than an ablation we reran. Bonferroni correction is applied to the 44-feature family at $\alpha = 0.05/44 = 0.0011$ but is reported alongside the asymmetric qualifying screen rather than serving as a substitute for it.

\begin{table}[!t]
    \centering
    \caption{H0 ablations for provenance and score-column redundancy. H0-minimal uses only external pre-score context and immutable timing. H0 no $\code{pos\_before}$ removes only the named focal-score column from the full H0 block. RF = random forest, LR = logistic.}
    \label{tab:pre-no-score}
    \begin{tabular}{llcccc}
    \toprule
    Split & Model & H0-minimal & H0 full & H0, no $\code{pos\_before}$ & $\Delta$ no-score \\
    \midrule
    Temporal $2024 \to 2025$ & RF & $0.6846$ & $0.7244$ & $0.7235$ & $-0.001$ \\
    Temporal $2024 \to 2025$ & LR & $0.6962$ & $0.7466$ & $0.7176$ & $-0.029$ \\
    GroupKFold(5)            & RF & $0.7447$ & $0.7814$ & $0.7705$ & $-0.011$ \\
    GroupKFold(5)            & LR & $0.7392$ & $0.7789$ & $0.7529$ & $-0.026$ \\
    \bottomrule
    \end{tabular}
\end{table}

\begin{table}[!t]
    \centering
    \small
    \caption{Minimum-detectable effect for the three structural features that are reported as flat in the joint specification, at $\alpha=0.05$ two-sided and power $0.80$.}
    \label{tab:null-triplet-mde}
    \setlength{\tabcolsep}{4pt}
    \begin{tabular}{lccccc}
    \toprule
    Feature & SE & Obs.\ OR/SD & Obs.\ $p$ & MDE OR/SD ($+$) & MDE OR/SD ($-$) \\
    \midrule
    $\code{rebuttal\_total\_len}$        & $0.0716$ & $1.02$ & $0.79$ & $1.22$ & $0.82$ \\
    $\code{n\_author\_replies}$          & $0.0637$ & $0.95$ & $0.44$ & $1.20$ & $0.84$ \\
    $\code{first\_reply\_latency\_days}$ & $0.0677$ & $0.90$ & $0.12$ & $1.21$ & $0.83$ \\
    \bottomrule
    \end{tabular}
\end{table}

\paragraph{Marginal univariate fits.} For each of the 44 feature IDs we additionally fit a logistic with that feature as the sole feature plus a constant, paper-clustered SE, on the same cohort. This produces a marginal OR for each feature, free of crowding from other feature indicators, used as one of the two gates in the asymmetric qualifying screen. Of the $22$ testable positive features (after dropping near-constant $F_{01}$), $22$ have marginal OR $> 1$. Of the $18$ testable negatives (after dropping the three rare negatives), $18$ have marginal OR $< 1$.

\paragraph{Polarity-isolated joint fit.} A second joint logistic keeps the headline H0 review-context, review-text, Gemini, tone, response-length, and author-reply controls, but restricts the feature block to negatives by dropping all 23 positive feature indicators. This tests whether negatives recover their declared direction once they no longer compete for variance with positives. Of $18$ testable negatives, $15$ have OR $< 1$ in this fit.

\paragraph{Robust association screen.} A feature passes the screen if its declared polarity is supported in BOTH (a) the marginal univariate fit AND (b) the polarity-isolated joint fit (headline joint fit for positives, negatives-only joint fit for negatives), with both 95\% CIs excluding $1$. The asymmetry between positives and negatives reflects the within-rebuttal multicollinearity documented above. Four of $22$ testable positive features (after dropping near-constant $F_{01}$) and ten of $18$ testable negative features (after dropping the three rare negatives at $<1\%$ prevalence) meet the screen.

\paragraph{Declared-polarity sign test.} As a single-number summary of polarity coherence we report a binomial test of "feature OR direction matches declared polarity" against the null hypothesis of a $50/50$ split, separately for positives and negatives. The headline joint fit gives positives $16/22$ ($p = 0.026$) and negatives $15/18$ ($p = 0.0038$). The no-structural ablation gives positives $15/22$ ($p = 0.067$) and negatives $14/18$ ($p = 0.015$). Stratum fixed effects give positives $14/22$ ($p = 0.143$) and negatives $14/18$ ($p = 0.015$). The marginal univariate fit gives positives $22/22$ ($p < 10^{-6}$) and negatives $18/18$ ($p < 10^{-5}$). The negatives-only joint gives negatives $15/18$ ($p = 0.0038$). Polarity is preserved on each side under every fit. The headline joint reports the conservatively crowded version of the positive block.

\paragraph{Robustness ablations.} The \code{No-structural-confounders} refit drops $\code{pos\_before}$, $\code{paper\_initial\_mean}$, $\code{paper\_initial\_std}$, $\code{n\_reviewers}$, $\code{confidence}$, the three sub-scores, and $\code{is\_initial\_outlier}$. The 44-feature ORs shift by a median $4.8\%$ in magnitude, polarity directions are preserved, and eight features cross Bonferroni in this ablation. The structural controls are therefore isolating signal rather than generating it. The \code{Stratum fixed effects} refit adds stratum dummies for $46$ ($\code{deviation\_quartile} \times \code{pos\_before\_quartile} \times \code{confidence\_quartile}$) cells (41 cells contain both outcome classes). Median absolute feature-OR shift vs the headline is $3.0\%$, and magnitudes are preserved within strata. This high-dimensional sparse-cell robustness fit reaches the iteration cap. The headline, marginal, no-structural, negatives-only, movability-stratum, and calibration fits converge. Diagnostics are written to $\code{statsmodels\_convergence.csv}$.

\begin{figure}[!b]
  \centering
  \includegraphics[width=0.7\textwidth]{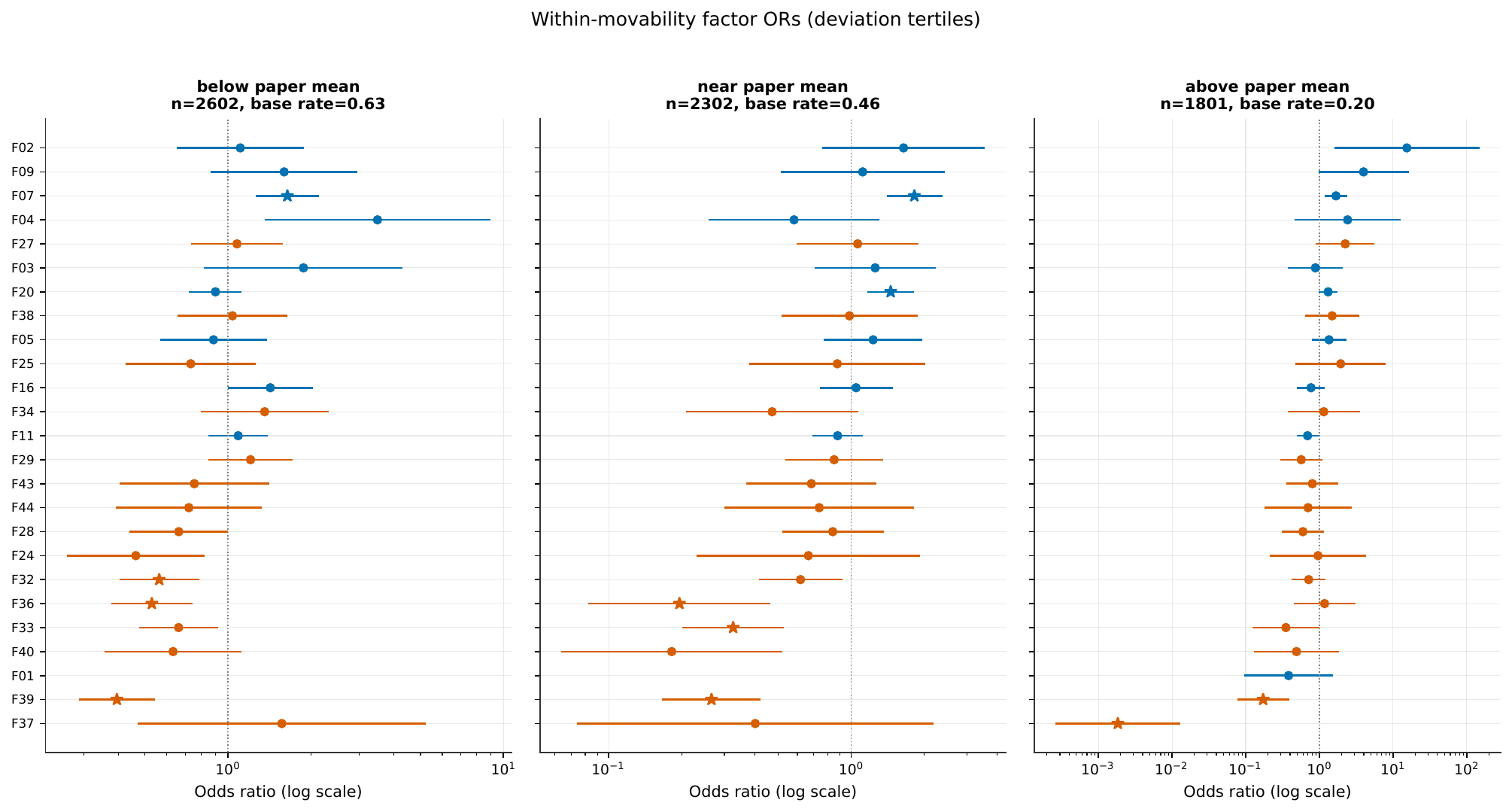}
  \caption{\textbf{Filtered stratum forest for resolved thread features.} Headline logistic refit on the cohort split by reviewer deviation from the paper-level initial mean: below ($n=2{,}602$), near ($n=2{,}302$), above ($n=1{,}801$). Markers show blue for positive, orange for negative, and $\star$ for crossing Bonferroni $\alpha=0.05/44$ within the stratum. The y-axis contains features with $|\log\text{OR}|>0.3$ in at least one stratum, so this is a visual diagnostic rather than a literal 44-row display. The full numerical feature-by-stratum table is in \apptab{tab:movability-feature-coefficients}.}
  \label{fig:movability-strata-app}
\end{figure}

\paragraph{H0-minimal and no-\code{pos\_before} ablations.} The leakage-sensitive H0-minimal ablation keeps only external pre-score context and immutable timing. It reaches temporal-split AUC $0.6962$ with the logistic model and $0.6846$ with the random forest (\apptab{tab:pre-no-score}). This confirms that a sizeable initial-review signal remains even when mutable OpenReview review-form text and Gemini-derived review-text measurements are excluded. The no-\code{pos\_before} refit answers a different question: whether the full H0 block is localised to the explicit $\code{pos\_before}$ column. Predictability barely moves for the random forest and modestly drops for the logistic. The caveat is that $\code{paper\_initial\_mean}$ and $\code{deviation\_from\_paper\_mean}$ jointly reconstruct $\code{pos\_before}$ via the identity $\code{pos\_before} = \code{paper\_initial\_mean} + \code{deviation\_from\_paper\_mean}$. We therefore treat the no-\code{pos\_before} fit as a redundancy check, while H0-minimal is the stricter provenance check.

\paragraph{Null-triplet minimum detectable effect.} The body reports rebuttal length, $\code{n\_author\_replies}$, and first-reply latency as flat in the joint specification. We document the power of that null. Cluster-robust SEs from the headline joint already account for paper-clustering, so the minimum detectable per-SD log-OR at significance level $\alpha = 0.05$ (two-sided) and power $1-\beta = 0.80$ is $\text{MDE}_{\log\text{OR}} = (z_{1-\alpha/2} + z_{0.80}) \cdot \text{SE} = 2.8016 \cdot \text{SE}$ (\apptab{tab:null-triplet-mde}). The detectable per-SD ORs for these three features are roughly $1.20$ to $1.22$ in the increase direction (or $0.82$ to $0.84$ in the decrease direction). The observed per-SD ORs are $1.02$, $0.95$, and $0.90$, all comfortably inside this noise envelope. The nulls therefore rule out moderate effects but not small ones.

\section{Statistical Framework}\label{app:auxiliary-stats}\label{app:stats}

Reviewer ratings are discrete and ordinal on a six-point scale with uneven spacing \{1, 3, 5, 6, 8, 10\} for 2023 to 2025. We use non-parametric tests throughout: Mann-Whitney $U$ with rank-biserial correlation for two-group comparisons, Kruskal-Wallis with epsilon-squared for multi-group, and Spearman $\rho$ rather than Pearson $r$ for correlations involving the score or confidence scales. For binary outcomes we use chi-squared or Fisher's exact tests with Cramer's V, and logistic regression with paper-clustered standard errors (statsmodels \code{cov\_type='cluster'}, grouped by \code{paper\_key}). Significance thresholds vary by analysis family: Bonferroni $\alpha = 0.05/44 \approx 0.0011$ for the 44-feature taxonomy validation; the asymmetric polarity screen of \Cref{sec:horizons} requires 95\% cluster-robust CI excluding $1$ in both marginal and polarity-isolated fits; structural covariates report at conventional $p < 0.05$ with cluster-robust CIs. With $n > \num{70000}$ trivial differences reach conventional significance, so we prioritise effect sizes over $p$-values. All claims use predictive language (``associated with'', ``predicts'', ``co-occurs with''). We do not claim causation. Reviews within a paper are not independent, so all prediction models use \texttt{GroupKFold} by \code{paper\_key} and all regression coefficients cluster on \code{paper\_key}.

\begin{longtable}{@{}L{0.65cm}L{0.7cm}L{5.2cm}R{2.1cm}R{2.1cm}R{2.1cm}@{}}
\caption{\textbf{Feature-level odds ratios within reviewer-paper movability strata.} Each cell reports OR [95\% CI] from a separate headline logistic refit within the below-, near-, or above-paper-mean stratum. \textemdash{} marks features dropped by the prevalence screen in that stratum. $^{*}$ marks Bonferroni significance at $\alpha=0.05/44$ within stratum.}
\label{tab:movability-feature-coefficients}\\
\toprule
ID & Pol. & feature label & Below mean & Near mean & Above mean \\
\midrule
\endfirsthead
\toprule
ID & Pol. & feature label & Below mean & Near mean & Above mean \\
\midrule
\endhead
\bottomrule
\endfoot
\bottomrule
\endlastfoot
F01 & pos. & Structured, point-by-point response format & \textemdash & \textemdash & $0.38$ [0.10, 1.53] \\
F02 & pos. & Non-defensive, respectful, collaborative tone & $1.11$ [0.65, 1.89] & $1.64$ [0.76, 3.56] & $15.58$ [1.59, 153.02] \\
F03 & pos. & Respectful opening / appreciation toward the reviewer & $1.88$ [0.82, 4.31] & $1.26$ [0.71, 2.23] & $0.89$ [0.37, 2.10] \\
F04 & pos. & The rebuttal signals sincere engagement through detailed responses, length, added appendix content, or extensive explanation & $3.49$ [1.36, 8.96] & $0.58$ [0.26, 1.31] & $2.41$ [0.46, 12.70] \\
F05 & pos. & Methodological detail and reproducibility clarification & $0.89$ [0.57, 1.39] & $1.23$ [0.77, 1.97] & $1.35$ [0.79, 2.33] \\
F06 & pos. & Concrete empirical evidence tied to reviewer concerns & $0.83$ [0.55, 1.24] & $0.87$ [0.58, 1.31] & $0.91$ [0.54, 1.53] \\
F07 & pos. & Baseline completion / comparator strengthening & $1.64$ [1.26, 2.15]$^{*}$ & $1.82$ [1.40, 2.37]$^{*}$ & $1.68$ [1.19, 2.38] \\
F08 & pos. & Demonstrating understanding and acknowledging valid points before responding & $1.21$ [0.86, 1.70] & $1.09$ [0.74, 1.59] & $1.14$ [0.64, 2.03] \\
F09 & pos. & Comprehensive concern coverage & $1.60$ [0.87, 2.96] & $1.12$ [0.51, 2.43] & $3.98$ [0.97, 16.28] \\
F10 & pos. & Explicit cross-references to figures, tables, and sections by number & $1.05$ [0.72, 1.54] & $0.95$ [0.67, 1.35] & $1.01$ [0.62, 1.65] \\
F11 & pos. & Ablation studies showing component-level impact & $1.09$ [0.85, 1.40] & $0.88$ [0.69, 1.12] & $0.69$ [0.49, 0.97] \\
F12 & pos. & (Proactive and) honest limitation acknowledgment & $0.89$ [0.68, 1.16] & $1.04$ [0.81, 1.32] & $0.77$ [0.56, 1.06] \\
F13 & pos. & Explicit manuscript revision signaling & $0.98$ [0.66, 1.45] & $0.99$ [0.65, 1.49] & $0.80$ [0.46, 1.41] \\
F14 & pos. & Supplementary material and new visual evidence & $1.03$ [0.76, 1.42] & $1.09$ [0.79, 1.49] & $0.94$ [0.61, 1.45] \\
F15 & pos. & Explaining novelty and broader impact & $0.79$ [0.55, 1.14] & $1.03$ [0.79, 1.34] & $1.04$ [0.73, 1.48] \\
F16 & pos. & Conceptual explanations alongside empirical results & $1.43$ [1.00, 2.04] & $1.05$ [0.74, 1.48] & $0.77$ [0.50, 1.19] \\
F17 & pos. & Clarification-oriented response strategy & $0.99$ [0.54, 1.83] & $0.96$ [0.54, 1.72] & $0.82$ [0.44, 1.52] \\
F18 & pos. & Concerns-addressed checklist or summary & $1.18$ [0.71, 1.97] & $0.92$ [0.55, 1.52] & $1.18$ [0.54, 2.56] \\
F19 & pos. & Rewriting and simplifying for clarity & $1.27$ [1.00, 1.61] & $1.27$ [1.01, 1.59] & $1.27$ [0.92, 1.76] \\
F20 & pos. & Theoretical or mathematical grounding & $0.90$ [0.72, 1.12] & $1.45$ [1.16, 1.82]$^{*}$ & $1.32$ [0.97, 1.78] \\
F21 & pos. & Active dialogue and continued engagement & $0.94$ [0.74, 1.20] & $1.17$ [0.92, 1.49] & $1.32$ [0.95, 1.85] \\
F22 & pos. & Concrete revision commitments & $1.08$ [0.71, 1.63] & $0.98$ [0.67, 1.43] & $1.12$ [0.68, 1.84] \\
F23 & pos. & Going beyond the explicit ask and anticipating follow-ups & $1.14$ [0.89, 1.45] & $1.02$ [0.81, 1.28] & $0.98$ [0.71, 1.36] \\
F24 & neg. & Defensive or dismissive tone & $0.46$ [0.26, 0.83] & $0.67$ [0.23, 1.92] & $0.96$ [0.21, 4.37] \\
F25 & neg. & Superficial or partial addressal of reviewer concerns & $0.73$ [0.42, 1.27] & $0.88$ [0.38, 2.03] & $1.95$ [0.47, 7.99] \\
F26 & neg. & Missing reproducibility details & $0.76$ [0.22, 2.65] & \textemdash & \textemdash \\
F27 & neg. & Missing or insufficient baseline comparisons & $1.08$ [0.74, 1.58] & $1.06$ [0.60, 1.90] & $2.24$ [0.90, 5.61] \\
F28 & neg. & Lack of quantitative/empirical evidence in rebuttal & $0.66$ [0.44, 1.00] & $0.84$ [0.52, 1.36] & $0.60$ [0.31, 1.15] \\
F29 & neg. & No new requested experiments / analyses / ablations / reviewer ``would-have-liked'' checks & $1.21$ [0.85, 1.71] & $0.85$ [0.53, 1.36] & $0.57$ [0.29, 1.10] \\
F30 & neg. & Vague future-oriented promises instead of concrete commitments & $1.17$ [0.86, 1.59] & $1.04$ [0.73, 1.48] & $0.90$ [0.55, 1.46] \\
F31 & neg. & Narrow or insufficient evaluation scope & $0.86$ [0.63, 1.19] & $0.81$ [0.53, 1.23] & $0.77$ [0.39, 1.49] \\
F32 & neg. & Incomplete empirical coverage even after rebuttal additions & $0.56$ [0.40, 0.79]$^{*}$ & $0.62$ [0.42, 0.92] & $0.72$ [0.42, 1.22] \\
F33 & neg. & Unresolved quality-of-execution concerns, methodological or experimental design issues that additional experiments do not fix & $0.66$ [0.48, 0.92] & $0.33$ [0.20, 0.53]$^{*}$ & $0.35$ [0.12, 1.01] \\
F34 & neg. & Silent acceptance of limitations with no mitigation & $1.36$ [0.80, 2.32] & $0.47$ [0.21, 1.07] & $1.14$ [0.37, 3.52] \\
F35 & neg. & Lack of structured rebuttal format, no explicit rebuttal section, no clear author-response markers, no point-by-point Q\&A with reviewer quotes & \textemdash & \textemdash & \textemdash \\
F36 & neg. & Weak novelty justification & $0.53$ [0.38, 0.74]$^{*}$ & $0.20$ [0.08, 0.47]$^{*}$ & $1.18$ [0.45, 3.10] \\
F37 & neg. & Brevity or low-effort rebuttal & $1.57$ [0.47, 5.22] & $0.40$ [0.07, 2.18] & $<0.01$ [$<0.01$, 0.01]$^{*}$ \\
F38 & neg. & Section-referencing or paper-rephrasing instead of real clarification & $1.04$ [0.66, 1.65] & $0.98$ [0.52, 1.88] & $1.49$ [0.64, 3.48] \\
F39 & neg. & Inadequate explanation or persistent clarity issues & $0.40$ [0.29, 0.54]$^{*}$ & $0.27$ [0.17, 0.42]$^{*}$ & $0.17$ [0.08, 0.39]$^{*}$ \\
F40 & neg. & Marginal quantitative improvements & $0.63$ [0.36, 1.12] & $0.18$ [0.06, 0.52] & $0.49$ [0.13, 1.85] \\
F41 & neg. & Presentation or notation overload that worsens clarity & \textemdash & \textemdash & \textemdash \\
F42 & neg. & Vague or insufficient technical explanation & $0.80$ [0.52, 1.22] & $0.97$ [0.49, 1.92] & $0.90$ [0.29, 2.82] \\
F43 & neg. & No explicit commitment to revise or improve the paper & $0.76$ [0.40, 1.41] & $0.68$ [0.37, 1.27] & $0.80$ [0.36, 1.81] \\
F44 & neg. & Surfacing a previously unnoticed problem without proactive framing & $0.72$ [0.39, 1.33] & $0.74$ [0.30, 1.82] & $0.71$ [0.18, 2.76] \\
\end{longtable}

\end{document}